# Synthesis and thermal stability of TMD thin films: A comprehensive XPS and Raman study

*Conor P. Cullen[1,2], Oliver Hartwig[3], Cormac Ó Coileáin [1,2], John B. McManus[1,2], Lisanne Peters[1,2], Cansu Ilhan[1,2], Georg S. Duesberg[1,2,3], Niall McEvoy[1,2]*

[1] School of Chemistry, Trinity College Dublin, Dublin 2, D02 PN40, Ireland

[2] AMBER Centre, CRANN Institute, Trinity College Dublin, Dublin 2, Ireland

[3] Institute of Physics, EIT 2, Faculty of Electrical Engineering and Information Technology, Universität der Bundeswehr, 85579 Neubiberg, Germany

## Abstract

Transition metal dichalcogenides (TMDs) have been a core constituent of 2D material research throughout the last decade. Over this time, research focus has progressively shifted from synthesis and fundamental investigations, to exploring their properties for applied research such as electrochemical applications and integration in electrical devices.

Due to the rapid pace of development, priority is often given to application-oriented aspects while careful characterisation and analysis of the TMD materials themselves is occasionally neglected. This can be particularly evident for characterisations involving X-ray photoelectron spectroscopy (XPS), where measurement, peak-fitting, and analysis can be complex and nuanced endeavours requiring specific expertise.

To improve the availability and accessibility of reference information, here we present a detailed peak-fitted XPS analysis of ten transition metal chalcogenides. The materials were synthesised as large-area thin-films on SiO$_2$ using direct chalcogenisation of pre-deposited metal films. Alongside XPS, the Raman spectra with several excitation wavelengths for each material are also provided. These complementary characterisation methods can provide a more complete understanding of the composition and quality of the material.

As material stability is a crucial factor when considering applications, the in-air thermal stability of the TMDs was investigated after several annealing points up to 400 °C. This delivers a trend of evolving XPS and Raman spectra for each material which improves interpretation of their spectra while also indicating their ambient thermal limits. This provides an accessible library and set of guidelines to characterise, compare, and discuss TMD XPS and Raman spectra.

## Introduction

Two-dimensional (2D) materials are at the forefront of materials research. In particular, transition metal dichalcogenides (TMDs) have been proposed for many future electronic,[1-3] photonic,[4,5] optoelectronic[6,7] or sensor devices.[8] High-performance devices based on TMDs have been demonstrated in virtually every field of nanotechnology, such as field effect transistors (FETs),[9] ultraviolet (UV) to infrared (IR) photodetectors,[10,11] micro- and nanoelectromechanical systems (MEMS/NEMS),[12,13] and gas sensors.[14,15] Moreover, newly discovered 2D materials and their heterostructures continue to reveal unique physical phenomena and observable quantum effects.[16-18]

The maturation of TMD synthesis procedures over the last ten years has naturally led to a shift from fundamental investigations of their properties towards studies focused on exploiting these properties for specific applications. This has encouraged research groups from a broad range of backgrounds to explore the utility of these materials in their respective areas of expertise.

The controlled large-area synthesis and the long-term stability of several TMDs remain critical issues. Consequently, comprehensive analysis of the morphology, chemical composition, and quality of the ultrathin films remains important. Often used characterization methods include scanning- and transmission electron microscopy (SEM, TEM) and scanning probe techniques for the morphology, while Raman spectroscopy and X-ray photoelectron spectroscopy (XPS) can reveal the elemental composition and chemical binding states. However, all these techniques need careful interpretation and analysis.

The analysis and peak fitting of XPS spectra for TMD materials is an often poorly implemented, under discussed, and yet essential part of a significant portion of 2D materials literature.[19] Assessments have shown ~30% of XPS analyses concerning next-generation materials are completely incorrect.[20] The reasons for this lack of focus are manifold and corrective action, in both this field and across wider XPS analysis, has been pleaded for by many of the world's foremost XPS experts.[19, 21] Recently, many educational works have been published to increase the accessibility and quality of XPS analysis.[22-27]

In this work, several of the most studied TMDs are synthesised as thin polycrystalline films by the thermally assisted conversion (TAC) of metal films. The films are thoroughly characterised using a combination of peak-fitted XPS and Raman spectroscopy with multiple excitation wavelengths. This includes the sulfides, selenides, and tellurides of Mo, W, and Pt. The primary purpose of gathering this characterisation data is to improve the availability of broadly applicable reference information, while using consistent methods which allow for internal comparison across the range of materials. Although characterisation of pristine TMD material establishes an important baseline reference, it is of potentially more limited utility as a reference point for interpreting more complicated spectra featuring imperfect materials encountered in common experimental conditions. Consequently, to expand the utility of the data in this work, the ambient thermal stability of the films is also assessed by annealing them at various temperatures in ambient conditions. The resulting various states of degradation and oxidation of the materials are characterised, compared, and discussed across the range of TMDs.

# Methods

**TMD film synthesis**

The synthesis methodologies for each material have previously been discussed in several publications,[28-36] a general method is provided here and the primary synthesis parameters given in Table.1.

Metal films were deposited on Si/SiO$_2$ substrates with controlled thickness using a Gatan precision etching and coating system (PECS 682). Thickness and deposition rate were closely monitored using a quartz-crystal monitor.

For the sulfide and selenide TMDs, substrates were placed in the primary heating zone of a two-zone quartz-tube furnace with a crucible filled with solid chalcogen placed in the secondary heating zone. The substrates were heated to their synthesis temperature under a forming gas environment (90% Ar/10% H$_2$, 150 sccm) with continuous vacuum pumping establishing a ~1 mbar atmosphere. When the substrates reached the synthesis temperature, the chalcogen was heated above its melting point for the desired reaction time. After this, the forming gas was changed to 100% Ar and the furnace was air cooled to room temperature before samples were retrieved.

A modified version of this procedure was used for PtS$_2$ synthesis, the substrates were placed into the centre of an open-ended quartz tube (17 mm diameter), and additional sulfur powder was placed at the entrance to this tube before it was placed into the primary heating zone of the furnace. Vacuum pumping was throttled during the synthesis to a pressure of ~200 mbar.[36]

The three telluride TMDs were synthesised by methods detailed in the literature.[30-32] This modified TAC process involved using a single-zone furnace for the synthesis process. A layer of Te was electrodeposited onto the transition metal film. This substrate was then placed inside a series of nested crucibles to maintain a higher partial pressure of Te during synthesis.

**Table.1** List of the studied TMDs and their synthesis parameters.

| Material | Chalcogen Precursor | Initial Metal Film (nm) | Carrier Gas | Primary Zone (°C) | Chalcogen Zone (°C) | Reaction Time (min) | Pressure (mbar) | Citation |
|---|---|---|---|---|---|---|---|---|
| **MoS$_2$** | S powder | 5 | Forming (90% Ar/10% H$_2$) | 750 | 130 | 25 | ~1 | [29] |
| **WS$_2$** | S powder | 5 | Forming | 800 | 130 | 25 | ~1 | [29] |
| **PtS$_2$** | S powder | 5 | Forming | 500 | 180 | 60 | ~200 | [36] |
| **PtS** | S powder | 5 | Forming | 500 | 130 | 60 | ~1 | [36] |
| **MoSe$_2$** | Se Pellets | 5 | Forming | 800 | 300 | 120 | ~1 | [35] |
| **WSe$_2$** | Se Pellets | 5 | Forming | 600 | 300 | 120 | ~1 | [35] |
| **PtSe$_2$** | Se Pellets | 5 | Forming | 450 | 300 | 120 | ~1 | [33] |
| **MoTe$_2$** | Te film | 20 | Nitrogen | 450 | N/A | 55 | ~700 | [30] |
| **WTe$_2$** | Te film | 20 | Nitrogen | 550 | N/A | 90 | ~700 | [32] |
| **PtTe$_2$** | Te film | 20 | Nitrogen | 450 | N/A | 180 | ~700 | [31] |

**Ambient annealing**

Each TMD film was split into 4 pieces, 3 of these were annealed on a VWR hotplate at one of 150, 300, and 400 °C for 30 minutes, in a controlled ambient environment at ~50% humidity.

This was paired with an external secondary thermocouple to verify the surface temperature of the hotplate.

**XPS measurements and analysis**

A PHI VersaProbe III system operated with SmartSoft-VersaProbe software was used for XPS measurements. This system uses a monochromated micro-focused scanning X-ray source with the Al $K\alpha$ X-ray line (1486.6 eV), under UHV pressures of approximately $10^{-10}$–$10^{-11}$ mbar. A dual-beam low energy electron flood gun and a low energy $Ar^+$ ion supply was automatically used to neutralise surface charges in the analysis region for each sample. Survey spectra were acquired with a pass energy of 112 eV and core-level spectra were generally acquired with a pass energy of 26 eV.

The energy range to acquire across for each core-level was determined by a combination of reference literature and an initial scan to determine approximately where a suitable background exists on either side of a spectrum. The most commonly changed parameter was the number of scans, and occasionally the pass energy (for the C 1s and O 1s only). These two parameters can be adjusted to optimise for spectral resolution and/or the scan time when attaining suitable signal to noise for a spectrum.

After measurement, all XPS data was imported into the software CasaXPS (version 2.3, Casa Software Ltd., U.K.) for analysis. Spectra were charge corrected using the C 1s binding energy value of 248.8 eV. After subtracting a Shirley-type background, core-level spectra were fitted with appropriate line shapes, generally a Gaussian-Lorentzian (e.g. GL(70) for a component with 70% Lorentzian and 30% Gaussian character), or for asymmetric features a Lorentzian asymmetric (LA) line shape was used ( i.e. LA($\alpha$, $\beta$, m) where $\alpha$ and $\beta$ define the spread of the peak on either side of the Lorentzian component and m is the width of the convoluting Gaussian component). Stoichiometry calculations were made by comparing the relative areas of the relevant components after accounting for their system-defined relative sensitivity factors. Fitted XPS spectra were exported and presented here using OriginPro (Version 2020b) software.

**Raman Measurements**

Raman spectroscopic analysis was performed using a 100x microscope objective (0.95 numerical aperture, spot size ~0.3 µm) with a WITec Alpha 300 R confocal Raman microscope. A 532, 405, or 633 nm excitation source at a power <200 µW and a spectral grating of 1800 lines/mm was used. All spectra were gathered by averaging measurements over >10 discrete point spectra along the surface.

**Plasma treatment**

A Diener PICO Barrel Asher (40 KHz, 100W, 0.3 mBar, 120 s) was used with $O_2$ gas to treat a 1.5 nm Pt metal film on $SiO_2$.

**Structural representations**

Atomic structural representations for each material were generated using VESTA 3 software.[37]

# Results and Discussion

**Characterisation considerations**

XPS characterisation of each film entailed measurement of their most relevant core-level regions, this varies per material but in general included, a TMD metal core-level (e.g. Mo 3d), a TMD chalcogen core-level (e.g. S 2p), the C 1s region, the O 1s region, and a survey spectrum. Alongside XPS, Raman spectra were measured with one of the most conventional excitation wavelengths 532 nm (2.33 eV) as well as with 633 nm and 405 nm (1.96 and 3.06 eV) excitations.

This process yielded a large amount of data due to each material being split into 4 samples as part of the thermal stability study. A minimum of 5 XPS spectra were acquired per sample, giving 20 XPS spectra per material, and thus 200 XPS spectra across the 10 different materials. The most useful spectra for this work are the TMD core-levels and so those will be the fitted spectra of interest. This reduces the number of essential spectra to 80 fully fitted, analysed, and compared XPS spectra. Alongside this, 3 Raman spectra were acquired per sample resulting in 120 Raman spectra of interest.

200 XPS and 120 Raman spectra of interest becomes an issue of data presentation and clarity. To manage this, the data will be broken into sections. First, the XPS fitting parameters and Raman of the unannealed "as-synthesised" material will be discussed for each individual TMD material, with the fitting details for the spectra given in the supplementary information. The thermal stability and oxidation results will then be discussed with comparison of the materials in each chalcogen grouping.

Charge correction and compensation is an essential aspect of XPS. The main impact that charge correction and charge referencing have on XPS spectra is on the energy position of the components. The most common methodology for charge correction is to shift the C-C ($sp^3$) component to ~285 eV, here the chosen value is 284.8 eV. This method is imperfect, but is widely accepted as the common practice.[38, 39] A potential issue with this method arises if there is poor electrical contact or inconsistent contact between the measured carbon material and the TMD material. This can give rise to significant errors for the quoted energy position of the components. This is not to say that the peak position numbers should be discounted entirely, but to acknowledge that there is potential error present and that a variation in the literature values for specific components can often be attributed to this.

Another aspect of XPS analysis of TMD materials lacking consistency is the energy window to measure and report for the specific core-levels. This can result in important information being missed or misinterpreted e.g. a common oxide being left outside the measured window and its contributions therefore being missed.

The presentation of peak-fitted XPS results is an important aspect of their potential utility. The high degree of user input in the results requires that the data is transparently presented to allow an informed assessment and critique of the peak-fitting. Beyond transparency in data analysis, the legibility and clarity of presented XPS data should be considered to ensure that the interpreted information is clearly conveyed and can be understood by those less familiar with the technique. To improve readability, spectra are presented here with background subtraction of the fitted components from the raw spectrum. To ensure the accuracy of the fitting implementation can still be determined, the envelope of the fitted peaks is not subtracted and is presented alongside the raw spectrum.

For Raman spectroscopy, consistent terminology regarding the symmetry, point groups, and Raman modes of the TMD materials as they are reduced in layer thickness towards monolayer can be troublesome, e.g. even and odd layer numbers of TMDs can possess different symmetries and therefore have different point groups and Raman modes. Common practice is to simply use bulk terminology for the material and its Raman modes unless specifically necessary to not. That convention is followed here for the thin film materials. Excitation energies in the region of 532 nm are the most common for TMD characterisation and so that excitation will be discussed most fully. The excitation wavelength chosen can influence the Raman cross section of different vibrational modes and any impact from this across both the TMD materials and their degradation products should be considered when drawing conclusions from Raman spectra, particularly when complementary characterisation methods are not used.

**General material properties**

Some of the most interesting features of the TMDs are their varying structures and properties, a number of these and some characteristic information for each material are given in Table.2. While there is not much structural diversity among the most common TMDs, with many adopting either the trigonal prismatic '2H' structure or the octahedral '1T' structure in their most stable forms, there are several less common structures, particularly for the telluride materials. It is possible to synthesise different structures directly, e.g. MoTe$_2$ is frequently cited as being synthesised in either the semiconducting 2H phase or the metallic 1T′ phase.[40, 41] As 1T' is the only MoTe$_2$ phase obtained here, the quoted information in Table.2 and the acquired spectra are specifically pertinent to that phase. Further structural information and representations are provided in Fig.S1-3

**Table.2** Material properties and characterisation features for the 10 materials studied here.

| Material | MoS$_2$ | WS$_2$ | PtS$_2$ | PtS | MoSe$_2$ | WSe$_2$ | PtSe$_2$ | MoTe$_2$ | WTe$_2$ | PtTe$_2$ |
|---|---|---|---|---|---|---|---|---|---|---|
| Bandgap (bulk/monolayer) eV | 1.3 / 1.9[42] | 1.4/ 2[43] | 0.25/ 1.8[44] | 1.36[45] | 1.1/ 1.6 [43] | 1.2/ 1.7[43] | Semimetal / 1.2 [44] | Semimetal[30] | Semimetal[46] | Metallic/ 0.9 [44, 47] |
| Structure | Trigonal Prismatic[42] | Trigonal Prismatic[43] | Octahedral[44] | Tetragonal[48] | Trigonal Prismatic[43] | Trigonal Prismatic[43] | Octahedral[44] | Monoclinic / Distorted Octahedral[30] | Orthorhombic/ Distorted Octahedral [46] | Octahedral[44] |
| Phase | 2H | 2H | 1T | - | 2H | 2H | 1T | 1T' | T$_d$ | 1T |
| Space group | P6$_3$/mmc | P6$_3$/mmc | P$\bar{3}$m1 | P4$_2$/mmc | P6$_3$/mmc | P6$_3$/mmc | P$\bar{3}$m1 | P2$_1$/m | Pmn2$_1$ | P$\bar{3}$m1 |
| Point group | D$_{6h}$ | D$_{6h}$ | D$_{3d}$ | D$_{4h}$ | D$_{6h}$ | D$_{6h}$ | D$_{3d}$ | C$_{2h}^2$ | C$_{2v}^7$ | D$_{3d}$ |
| Main Raman Modes | A$_{1g}$, E$_{1g}$, E$_{2g}^1$ | A$_{1g}$, E$_{1g}$, E$_{2g}$ | A$_{1g}^1$, A$_{1g}^2$, E$_g$ | B$_g$ | A$_{1g}$, E$_{1g}$, E$_{2g}$ | A$_{1g}$, E$_{1g}$, E$_{2g}$ | A$_{1g}$, E$_g$ | A$_g$, B$_g$ | A$_1$, A$_2$, B$_1$, B$_2$ | A$_{1g}$, E$_g$ |
| XPS regions (binding energy eV) | *Mo 3d* 222-240, *S 2p* 158-173 | *W 4f* 30- 45, *S 2p* 158- 173 | *Pt 4f* 68-82, *S 2p* 158-173 | *Pt 4f* 68-82, *S 2p* 158-173 | *Mo 3d* 224-240, *Se 3d* 51-62 | *W 4f* 30-45, *Se 3d* 51-62 | *Pt 4f* 68-82, *Se 3d* 51-62 | *Mo 3d* 224-240, *Te 3d* 569-591 | *W 4f* 30-48, *Te 3d* 569-591 | *Pt 4f* 68-82, *Te 3d* 569-591 |

**MoS₂**

Molybdenum disulfide is by a significant margin the most studied TMD. Both the XPS and Raman spectra of MoS₂ are widely reported in the literature.

*XPS*

The most frequently studied and reported XPS core-level regions for MoS₂ are the Mo 3d (~230 eV) and S 2p (~160 eV). The Mo 3d region has a coincidental overlap with the S 2s core-level, this can lead to potential confusion and incorrect fitting.

An appropriate measurement region for the Mo 3d, to include both the commonly occurring Mo 3d components and those from the S 2s, is determined to be ~222–240 eV. The XPS spectra of the as-synthesised material in this study in this energy region are shown in Fig.5.3(a). There are three distinct states fitted in this region, consistent with literature.[49, 50]

The major features will be discussed, going from the low to high binding energy side of the figure. The first is S 2s at 226.1 eV, this is fitted most accurately with an SGL(60) line shape, a sum Gaussian-Lorentzian line shape, this shows increased Lorentz-like features with a reduced peak height. The most important component is the MoS₂ Mo 3d doublet, with a peak separation of 3.14 eV, and the $Mo\ 3d_{5/2}$ at 228.9 eV. The full width at half maximum (FWHM) of 0.83 eV is an indicator of the high crystallinity and therefore high uniformity in the environment of the $Mo^{4+}$ atoms in the MoS₂ and can be used as a comparative baseline to indicate crystallinity or disorder. It should be noted however that the measured peak width is subject to experimental parameters such as the pass energy and so is only directly comparable with this dataset.

The $Mo\ 3d_{5/2}$ of the most common stable oxide, MoO₃, is at 232.3 eV and therefore overlaps with the MoS₂ $Mo\ 3d_{3/2}$, the presence of MoO₃ can be immediately observed before fitting if the measured region includes the MoO₃ $Mo\ 3d_{3/2}$ at ~235.5 eV, not including this can lead to misinterpretation, misrepresentation, or inadequate fits.

The chalcogen core-level used for MoS₂ characterisation is the S 2p. Unlike for the Mo 3d region, there is generally no overlap between the TMD and the common oxide components for sulfur. This can regularly result in the reported XPS region being too narrow to confirm the purity of the material as oxides are outside the measured region, the measured region here is ~158–173 eV. While it is very common to see low levels of molybdenum oxides in MoS₂, the same is not true for trace sulfur oxides. There is only one doublet detected in this region corresponding to MoS₂, with the $S\ 2p_{3/2}$ at 161.8 eV with a peak separation between the $S\ 2p_{3/2}$ and $S\ 2p_{1/2}$ of 1.16 eV.

*Raman*

Raman spectroscopy of MoS₂ is a well-developed area of the literature. As with most TMDs, a ~532 nm excitation wavelength is likely the most common for characterisation of MoS₂. Alongside this, 405 and 633 nm lasers are also used here.

The typical Raman spectra for MoS₂ with the different excitations are shown in Fig. 1(c,d,e), for 532 nm the characteristic modes of MoS₂ are the $E_{2g}^1$ and the $A_{1g}$ at ~ 384 and 410 rel cm⁻¹ respectively. The Raman signal here is characteristic of bulk-like MoS₂ as previously seen for similar films.[51] Another common feature of the MoS₂ Raman spectrum is the broad component at ~ 460 rel cm⁻¹ this is attributed to phonon interactions and is assigned as $2LA(M)$, as it involves 2 longitudinal acoustic phonons from the edge of the Brillouin zone at the M point.[51] Similar to XPS, the width of the Raman peaks can be indicative of the crystallinity and uniformity of the structure.[52, 53]

Excitation at 405 nm shows very similar features to the 532 nm, with the $E_{2g}^1$ and $A_{1g}$ peaks being prominent. There is also an inversion in relative intensity with the $A_{1g}$ now being lower in intensity than the $E_{2g}^1$ is due to a resonance effect at 532 nm.[54]

Excitation at 633 nm (1.96 eV) provokes a substantially different spectrum. This is due to wavelengths in this region generating a resonant Raman spectrum from MoS₂ resulting in additional peaks. The energy proximity to the direct bandgap of MoS₂ for this excitation wavelength is the cause of this resonance. The relative lowering of intensity attributed to the first-order Raman modes ($E_{2g}^1$ and $A_{1g}$) and significant increase in second-order Raman features is a well known occurrence.[55, 56]

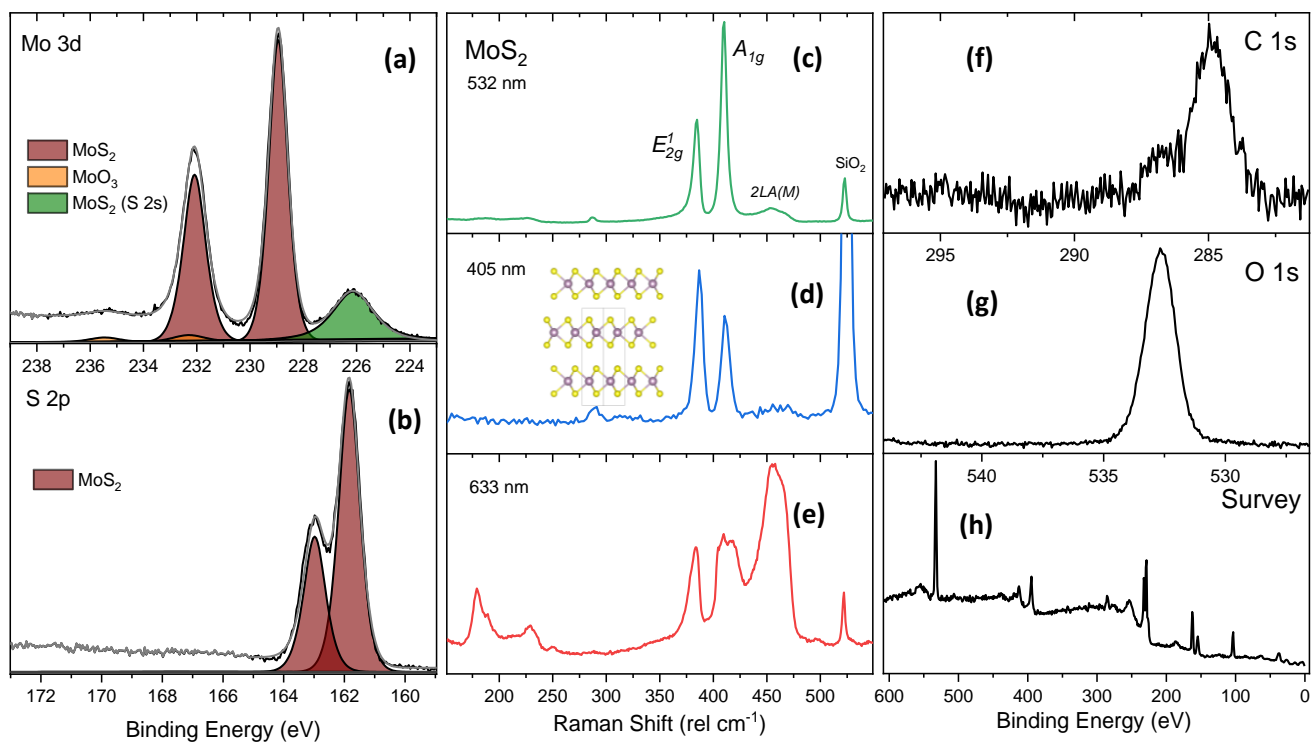

**Fig.1** Peak-fitted XPS spectra of the Mo 3d **(a)** and S 2p **(b)** core-levels. Raman spectra of MoS$_2$ acquired with 532 nm **(c)** 405 nm **(d)** and 633 nm **(e)** excitation energies with insert of the MoS$_2$ structure. XPS spectra of the C 1s **(f)**, O 1s **(g)** and survey **(h)** regions.

## WS$_2$

The regions of interest for XPS analysis of WS$_2$ characterisation are the W 4f (~33 eV) and the S 2p (~160 eV) and the resultant data is presented in Fig. 2.

### *XPS*

The W 4f core-level region is the most studied W level in XPS characterisation due to the relative intensity of the core-level and the significant shifts in binding energy position with chemical environment. An appropriate energy window to measure the W 4f is determined here to be ~30–45 eV. It has a coincidental overlap with the $W\ 5p_{3/2}$, the splitting of this core-level is large enough that only the $5p_{3/2}$ is located in this window. Fitting and accurately accounting for the $5p_{3/2}$ components in complex W XPS spectra can be difficult and a significant source of error or confusion if poorly accounted for.

As an example of one of the many issues prevalent in TMD XPS literature, the WS$_2$ $W\ 5p_{3/2}$ in oxide-free material can be potentially mistaken for a W$^{6+}$ or WO$_3$ W 4f component, giving an inaccurately negative assessment of material purity.[57-61] Inversely, the same can be true of W 4f WO$_3$ components being instead attributed to the WS$_2$ $W\ 5p_{3/2}$, giving an inaccurately positive assessment of material purity.[62-65]

Here three separate components are measured in the W 4f region, all of them fitted with a GL(60) peak shape. These are a W 4f doublet corresponding to WS$_2$ with the $W\ 4f_{7/2}$ at 32.5 eV and a peak splitting of 2.18 eV, and the corresponding WS$_2$ $5p_{3/2}$ at 38.2 eV. There is a very low level of oxide present in the form of WO$_3$, with its $W\ 4f_{7/2}$ at ~36 eV. The WO$_3$ also has its corresponding $5p_{3/2}$ component, but this is below the detection limit here.

Similar to MoS$_2$, the S 2p core-level region is used for characterisation. The same practice of using a wide energy window (~158–173 eV), to ensure that any oxides present are detected, is followed here. As was also the case for MoS$_2$, there are no other detected components besides the WS$_2$ S 2p, with the $S\ 2p_{3/2}$ at 162 eV.

*Raman*

The Raman spectra of WS$_2$ with the three excitation wavelengths are shown in Fig.2(c–e). WS$_2$ has the same symmetry as MoS$_2$ and similarly, the $E^1_{2g}$ and $A_{1g}$ Raman modes are the most prominent for WS$_2$.[66] 532 nm Raman spectroscopy of WS$_2$ is significantly more complex than MoS$_2$ as it has a high number of second-order peaks, the most significant is the $2LA(M)$ which overlaps heavily with the $E^1_{2g}$ resulting in a broad double component feature at ~350 rel cm$^{-1}$. For few-layer WS$_2$, the ratio of $A_{1g}$ to $2LA(M)$ can be used to identify monolayers, with an exceptionally pronounced $2LA(M)$ component present in monolayer WS$_2$ due to resonance effects.[67]

The Raman spectra of the WS$_2$ film acquired using a 405 nm excitation laser shows some clear differences from that acquired with the 532 nm. There are markedly fewer second-order peaks as this excitation is off resonance, this is most pronounced in the significantly lower contribution of the $2LA(M)$ resulting in a mostly separated $E^1_{2g}$ at ~355 rel cm$^{-1}$. For 405 nm excitation there is also an apparent feature at ~155 rel cm$^{-1}$ this is not a Raman feature and is simply a result of the spectrometer cut-off and is present in all Raman spectra gathered with 405 nm excitation here.

The 633 nm excitation spectrum shows a similar number of second-order components to the 532 nm. The most unique feature in this spectrum is multiple peak nature of the $A_{1g}$ at ~420 rel cm$^{-1}$, unlike the single peak structure for the other wavelengths. This has been seen in cases where the excitation energy is close in energy to the A exciton energy in WS$_2$, it has been proposed this multi-peak structure is a result of the $A_{1g}$ component from the distinct layers giving varying frequency Raman components. [68]

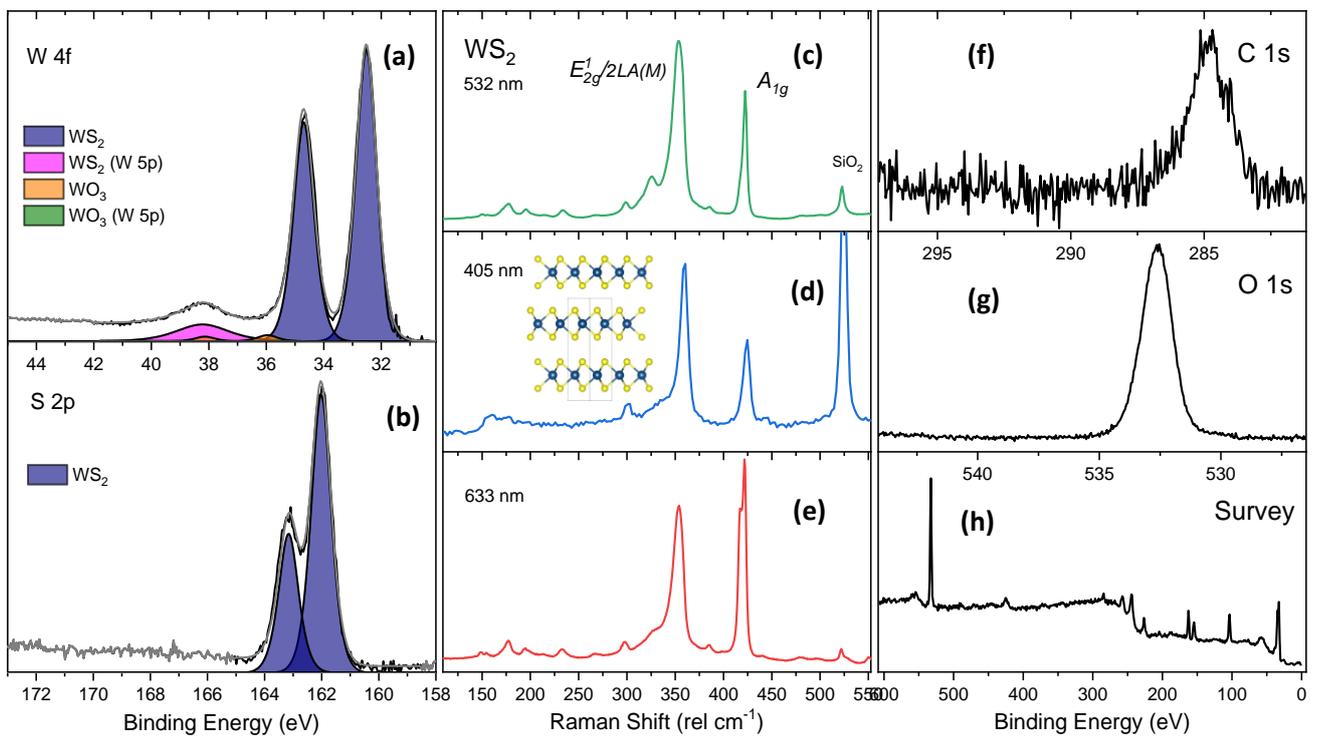

**Fig. 2** Peak-fitted XPS spectra of the W 4f **(a)** and S 2p **(b)** core-levels. Raman spectra of WS$_2$ acquired with 532 nm **(c)** 405 nm **(d)** and 633 nm **(e)** excitation energies with insert of the WS$_2$ structure. XPS spectra of the C 1s **(f)**, O 1s **(g)** and survey **(h)** regions.

**PtS$_2$**

To date, literature studies of PtS$_2$ have been relatively sparse. This is in part due to the difficulty associated with synthesising PtS$_2$ films. One of the complications in the synthesis and the characterisation of PtS$_2$ is the higher relative stability of the non-layered monosulfide PtS, which forms more readily under standard TMD sulfide synthesis conditions.[36, 69]

As a result of the scarcity of research into this material, XPS and Raman spectra of thin films of PtS$_2$ have only recently been discussed in detail.[34, 36, 69-72]

*XPS*

XPS of PtS$_2$ focuses on the regions of the Pt 4f (~72 eV) and the S 2p (~160 eV), the resultant data is shown in Fig. 3.

The energy window for measurements of the Pt 4f core-level here is 68–82 eV. There are three chemical states detected in the Pt 4f region. Going from low to high binding energy, the first doublet with $Pt\ 4f_{7/2}$ at 72.1 eV is attributed to Pt$^{2+}$ in the form of PtS, it is found to be a common contaminant in PtS$_2$. This component can be potentially misattributed to Pt metal due to the similar expected peak positions. The symmetric line shape of this component is a clear sign against this, as Pt metal has an asymmetric XPS line shape.

The main PtS$_2$ $Pt\ 4f_{7/2}$ component is found at 73.8 eV in this work, with a peak splitting for the Pt 4f doublet of 3.33 eV.

A third feature is fitted at 75.3 eV, this has been tentatively attributed to a Pt oxide, most likely PtO$_2$.[73] The exact fitting and peak identification of this component is up for debate due to its consistently low area and significant overlap with both of the other $Pt\ 4f$ states fitted here. Evidence indicating the high probability of this being PtO$_2$ will be explored in more detail in the PtSe$_2$ section.

To measure any likely oxides, the energy window for measuring the S 2p region is ~158–173 eV for PtS$_2$. The S 2p region of PtS$_2$ is also markedly underexplored in the literature and is significantly more complex than those seen for the other TMD sulfides.

Four separate chemical states are identified here during deconvolution in this region. Going from low to high binding energy they are the $S\ 2p_{3/2}$ for PtS$_2$ at 162.2 eV, then the $S\ 2p_{3/2}$ for the low level of PtS at 162.6 eV.

Unlike the other sulfide TMDs, low levels of elemental sulfur are identified at $S\ 2p_{3/2}$ 163.8 eV. Interestingly, a broad feature at 166.7 eV is observed here and is attributed tentatively to a S$^{4+}$ component, likely from a sulfite species.[74] This consistently broad feature has been observed previously for platinum sulfide materials albeit with little or no discussion of its origin,[75-77] it is currently unexplained and may not be an oxide component. The catalytic properties of Pt and its interactions with sulfur have traditionally been studied with respect to catalytic convertors. The disproportionation of SO$_2$ to a variety of sulfates, sulfites, and sulfides in different environments and on different substrates has been discussed in this context.[78, 79] While these systems are substantially different from our experimental system, a similar role of Pt atoms during the TMD synthesis process particularly concerning its interactions with SO$_2$ are worth considering for future study and may contribute to the unique aspects of its XPS spectrum.

*Raman*

The Raman spectra of PtS$_2$ are shown in Fig.3(c)-(e).

PtS$_2$ has an octahedral or 1T structure and a $D_{3d}$ point group.[34, 72] For 532 nm excitation there are three intense Raman modes from PtS$_2$, the most intense being the $E_g^1$ at 305 rel cm$^{-1}$. The second main PtS$_2$ mode is the $A_{1g}^2$ at 344 rel cm$^{-1}$ this has ~40% the intensity of the $E_g^1$ mode in these films.

Some complications arise due to the third component, the $A_{1g}^1$ at 336 rel cm$^{-1}$. From the XPS analysis it is known there exists a low level of PtS in the measured material, PtS has its main Raman component at 335 rel cm$^{-1}$ which tightly overlaps with the $A_{1g}^1$ of PtS$_2$, this is a source of some potential error but may also serve as a method to identify PtS contamination in the material via Raman spectroscopy.

Additionally, there is a shoulder to this 532 nm Raman spectra at ~360 rel cm$^{-1}$, it is currently unassigned in literature. Using a 405 nm Raman laser yields a significantly lower intensity Raman signal than for 532 nm, it is difficult to ascertain any changes in features to the spectrum beyond this. The lowering of intensity and broadening of the components makes any assignment of the heavily overlapping peaks less accurate.

The 633 nm Raman spectrum has a high intensity and shows significantly narrower Raman modes than for the previous wavelengths. This shows the three previously discussed components clearly, the separation of the $E_g^1$ and the $A_{1g}^1$ also narrows from 39.7 to 34 rel cm$^{-1}$ when comparing the 532 and 633 nm Raman spectra respectively. This positions 633 nm excitation as potentially the most beneficial excitation wavelength considered here for PtS$_2$ characterisation.

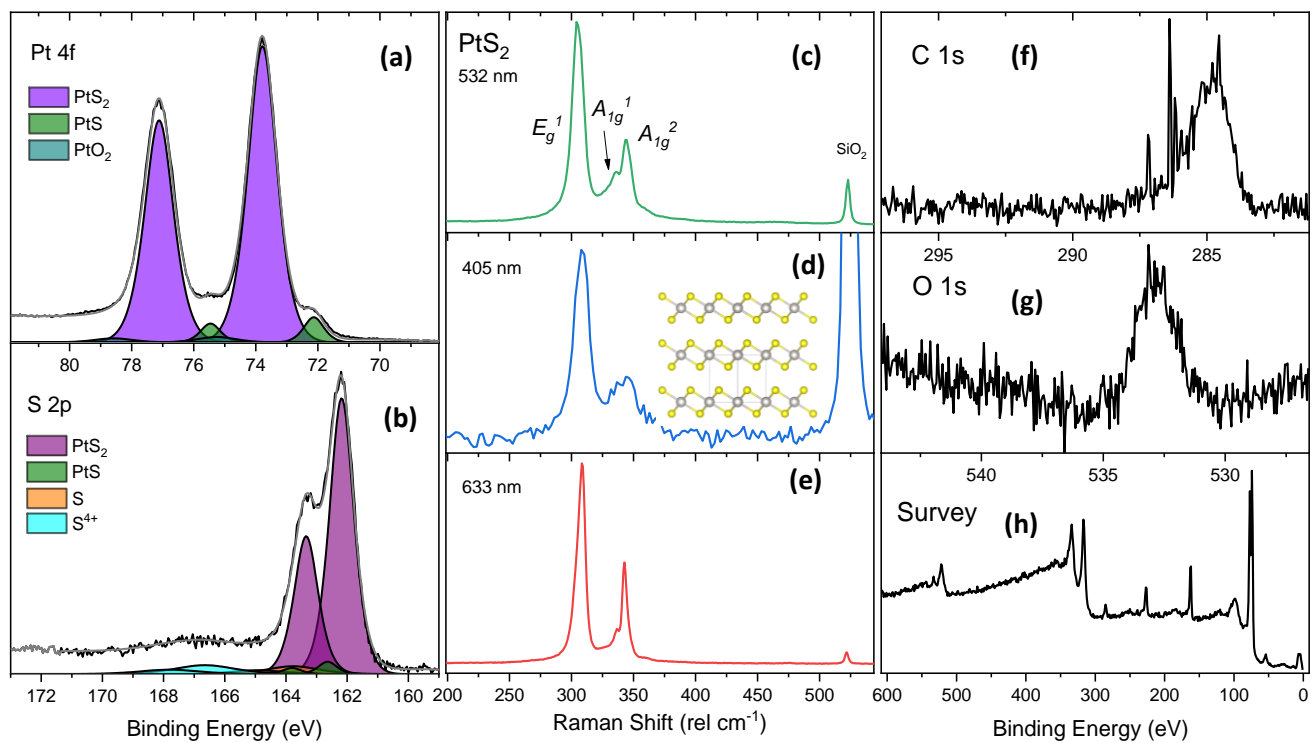

**Fig. 3** Peak-fitted XPS spectra of the Pt 4f **(a)** and S 2p **(b)** core-levels. Raman spectra of $PtS_2$ acquired with 532 nm **(c)** 405 nm **(d)** and 633 nm **(e)** excitation energies with insert of the $PtS_2$ structure. XPS spectra of the C 1s **(f)**, O 1s **(g)** and survey **(h)** regions.

**PtS**

The $PtS/PtS_2$ system is unique among the materials analysed in this work as it the only TMD which preferentially forms a mono-chalcogenide over a range of common synthesis conditions.

PtS is not a layered material; it has a tetragonal structure as shown in Fig.S1 it is therefore markedly different than any of the other materials looked at here. The importance of PtS in the characterisation of $PtS_2$, and the limited amount of characterisation literature of PtS in general elevates the necessity to include PtS here.[45] PtS, known as a constituent of the mineral Cooperite, is more commonly examined in the realm of geology than nanomaterials and so characterisation of PtS with the techniques common in 2D materials is developing.[36, 45, 77]

*XPS*

XPS of PtS follows the same practice and parameters as that of PtS$_2$ focusing on the regions of the Pt 4f (~72 eV) and the S 2p (~160 eV). The resultant data is presented in Fig. 4.

The Pt 4f core-level is measured between ~68–82 eV. Two sets of Pt 4f doublets are fitted in this spectrum with a dominant $Pt\ 4f_{7/2}$ PtS peak at 72.1 eV.

A high-energy shoulder is fitted with the $Pt\ 4f_{7/2}$ at 73.3 eV and is attributed to contamination with PtS$_2$. The low intensity of this component and its overlap with the PtS components can make it difficult to assign. However, by quantitative comparison with the S 2p spectrum, the relative area of the PtS$_2$ $S\ 2p_{3/2}$ can be used to more accurately fit and determine that the high-energy shoulder in the Pt 4f spectrum is most likely PtS$_2$ and not an oxide or other feature.

Similar to PtS$_2$, the chosen energy window for the S 2p core-level is ~158–172 eV. The PtS spectrum measured here is noticeably similar, in terms of components, to the PtS$_2$ S 2p spectrum shown in Fig.3. There are four distinct chemical states fitted in the spectrum. Going from low to high binding energy, the first is a small PtS$_2$ component with $S\ 2p_{3/2}$ at 161.3 eV. Next, the main PtS $S\ 2p_{3/2}$ is fitted at 162.5 eV with a FWHM of 0.81 eV. The final two states are a substantial elemental sulfur component fitted at 164 eV and a broad feature with $S\ 2p_{3/2}$ at 166.8 eV which is provisionally assigned to a S$^{4+}$ species, likely in the form of a sulfite as discussed for PtS$_2$.

*Raman*

Raman spectroscopy of thin-film PtS is a scarcely studied area in the literature. As a result of this, the complex aspects of these Raman spectra are not well understood, and an in-depth theoretical study would be required in order to determine the nature of all aspects of the spectra.[45]

Looking first to the 532 nm Raman of PtS in Fig.4 (c), the largest peak in the spectrum is from the SiO$_2$ substrate. This is indicative of the significantly lower Raman signal generated by PtS relative to the TMDs measured previously, all of which gave much stronger Raman signal for similar films.

The most prominent Raman component, and the only commonly reported component of the PtS Raman spectrum, is the major feature at ~330 rel cm$^{-1}$. This is a combination of two modes, the majority contribution is the $B_{1g}$ mode at 335 rel cm$^{-1}$ with a shoulder from the $E_g$ at ~326 rel cm$^{-1}$. [45]

Approximately 10 low-intensity features are spread throughout the spectrum. These are likely to be from a combination of first-order and second-order Raman features, due to the limited literature in this area, these are currently unassigned.

The 405 nm Raman spectrum of PtS yields very little information due to the very low intensity Raman signal at this excitation. The expected main component is clearly present at 334 rel cm$^{-1}$. The feature at ~150 rel cm$^{-1}$ is due to the spectrometer cut off and is not a PtS component.

Interestingly the Raman spectrum of PtS using 633 nm excitation wavelength yields a substantially different spectrum than the other two wavelengths. Unlike the previous spectra, there are at least four prominent Raman peaks from PtS in this spectrum. The $B_{1g}$ mode at 334 rel cm$^{-1}$ and three other currently unreported and unassigned modes at 115 rel cm$^{-1}$, 375 rel cm$^{-1}$ and 475 rel cm$^{-1}$. Many of the unassigned low-intensity features observed in the 532 nm excitation spectrum are detectable in this spectrum also. The presence of only some of these peaks in both spectra is possibly due to resonance effects caused by the different wavelengths.

This is congruent with the Raman findings for PtS$_2$ and further exemplifies the utility of using 633 nm excitation when examining the platinum sulfide materials with Raman spectroscopy.

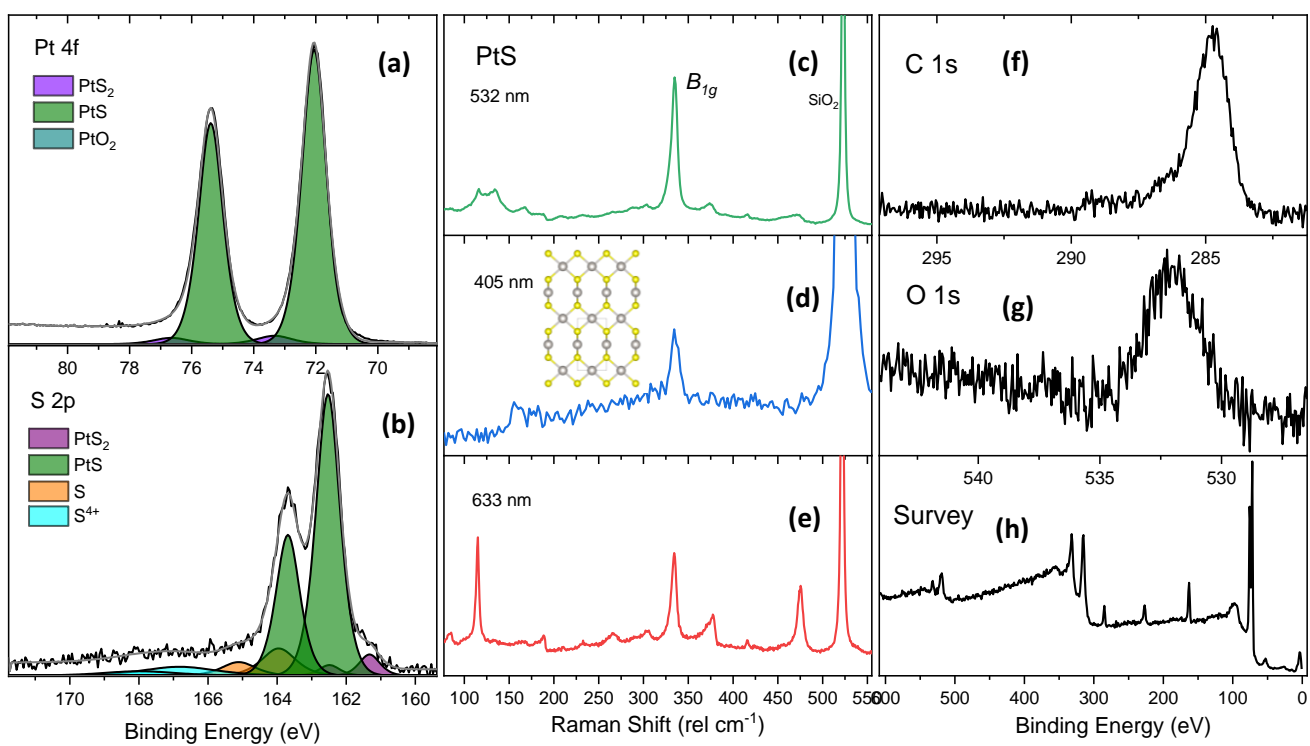

**Fig 4** Peak-fitted XPS spectra of the Pt 4f **(a)** and S 2p **(b)** core-levels. Raman spectra of PtS acquired with 532 nm **(c)** 405 nm **(d)** and 633 nm **(e)** excitation energies with insert of the PtS structure. XPS spectra of the C 1s **(f)**, O 1s **(g)** and survey **(h)** regions.

**TMD Selenides**

**MoSe₂**

*XPS*

The XPS regions of interest for MoSe$_2$ are the Mo 3d (~230 eV) and the Se 3d (~55 eV), and the resultant data is shown in Fig.5.

The energy window for measurement of the Mo 3d is 224–240 eV here. Three components are fitted in this spectrum. The main Mo 3d doublet has the $Mo\ 3d_{5/2}$ at 228.6 eV, which has a spin-orbit splitting of 3.14 eV for the $Mo\ 3d_{3/2}$. A significant complication in the MoSe$_2$ Mo 3d spectrum is the presence of the coincidentally overlapping Se 3s core-level.[50, 80-83] Here the

Se 3s is at 229.4 eV. The Se 3s overlaps with both MoSe$_2$ Mo 3d components which can easily lead to confusion or inaccurate fitting. The Se 3s for MoSe$_2$ is frequently missing from XPS analysis in the literature, resulting in fitting and interpretation errors.[84-88]

The Se 3d is measured between 51–62 eV, shown in Fig. 5(b). The Se 3d core-level is fitted here with two doublets, they are the main MoSe$_2$ doublet with $Se\ 3d_{5/2}$ at 54.3 eV, and a high binding energy shoulder is resolved by including a second doublet with $Se\ 3d_{5/2}$ at 54.6 eV. The spin-orbit splitting of the Se 3d being 0.86 eV results in significant overlap of these components, opening the possibility for error in fitting.

This second doublet seen in Fig. 5(b) is attributed to a sub-stoichiometric MoSe$_x$ species, this is difficult to unambiguously assign and could potentially be from an undetermined SeO$_x$ species. To be qualitatively consistent, there should be a corresponding Mo 3d MoSe$_x$ feature, which is not observed here. The lack of a corresponding component may be explained by the Mo 3d features for MoSe$_2$ and MoSe$_x$ overlapping to a high degree. In that scenario it would then be expected to see the "MoSe$_2$" Mo 3d peaks fitted with significantly wider FWHM than its corresponding MoSe$_2$ Se 3d component, that is the case here (0.92 and 0.6 eV) lending credence to the assignment of the secondary Se 3d species as MoSe$_x$. A similar overlapping high binding energy shoulder is found for all 3 selenide materials measured here; it is also frequently seen in the literature.[28, 89] The peak position of this extra component varies for each TMD material. This, within error, makes the assignment of oxide unlikely as there should be negligible change in SeO$_x$ peak position between the different TMDs. Further indication comes from quantitative analysis, as not including this sub-stoichiometric TMD component when calculating the stoichiometry of the selenide TMD materials yields significantly divergent results compared to the other TMDs. Together these arguments leave a MoSe$_x$ species as the most probable assignment.

A further complicating feature of selenide TMD XPS is the overlap of the C 1s core-level (~285 eV) and the Se Auger peaks at ~ 287 and 299 eV, as seen in Fig.5(f). As charge referencing is used against the C 1s C-C at 284.8 eV, there is significant potential for increased peak position error. This is one of the many potential issues with reliance on C 1s charge correction.

*Raman*

Raman spectroscopy of MoSe$_2$ is a mature area of research with well resolved spectra reported from at least 1980.[90]

MoSe$_2$ has the same structure as MoS$_2$, that is the 2H trigonal prismatic structure with point group D$_{6H}$. The 532 nm Raman spectrum of MoSe$_2$ has four prominent modes, the $E_{1g}$, $A_{1g}$, $E_{2g}^1$, $A_{2u}$ at 170, 244, 286, 352 rel cm$^{-1}$ respectively. There are several other modes of lower intensity, discussion of these is left to previously published works in the literature.[91, 92]

The 633 nm excitation Raman spectrum of MoSe$_2$ is similar to the 532 nm spectrum, the most notable changes being the lower relative signal from the $E_{1g}$ and $A_{2u}$ peaks.

The 405 nm excitation Raman spectrum of MoSe$_2$ is complex and noticeably different from the 532 nm spectrum. The most prominent change is the dramatically lower intensity of the $A_{1g}$ mode and the increase of the previously diminutive $E_{2g}^1$. It has previously been postulated this is likely due to a resonance effect with the C exciton of MoSe$_2$.[91, 92]

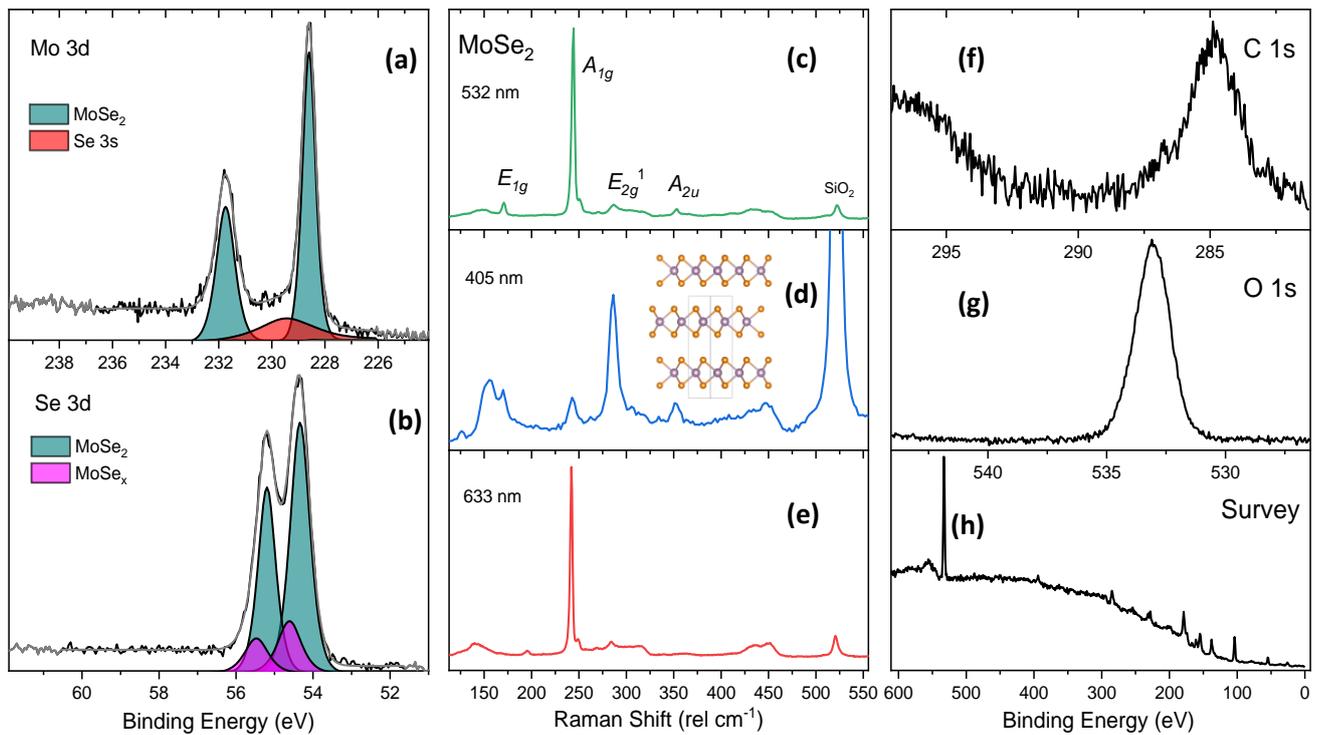

**Fig 5** Peak-fitted XPS spectra of the Mo 3d **(a)** and Se 3d **(b)** core-levels. Raman spectra of MoSe$_2$ acquired with 532 nm **(c)** 405 nm **(d)** and 633 nm **(e)** excitation energies with insert of the MoSe$_2$ structure. XPS spectra of the C 1s **(f)**, O 1s **(g)** and survey **(h)** regions.

## WSe₂

WSe₂ is the most widely studied of the selenide TMDs in the literature, likely due to its typically p-type or ambipolar conductivity in contrast to mostly n-type for other common TMDs.[93]

*XPS*

The XPS regions of interest are the W 4f (~33 eV) and the Se 3d (~55 eV) and their resultant data is shown in Fig.6.

As was the case for WS₂, the energy window for the $W\ 4f$ is ~30–45 eV. WSe₂ is fitted here with six components from only two chemical states of W. The main WSe₂ $W\ 4f_{7/2}$ is fitted at 32.6 eV with a 2.18 eV spin-orbit splitting for the $W\ 4f_{5/2}$. The corresponding $W\ 5p$ component is fitted at 35.4 eV.

There is a significant amount of WO₃ measured in this spectrum, ~40% of W atoms, with the $W\ 4f_{7/2}$ of WO₃ fitted at 36.1 eV with a corresponding $W\ 5p$ at 41.8 eV. Due to the complete overlap of the WO₃ $W\ 4f$ components and the WSe₂ $W\ 5p$, it can be difficult to accurately account for all the individual components in this spectrum. Nevertheless, by fixing the ratio of $W\ 4f_{7/2}$ area to $W\ 5p$ area, and finding a fixed binding energy separation, these overlapping components can be accounted for consistently. This required referencing a $W\ 4f$ spectra with little oxide, in this work the WS₂ examined previously was chosen. Accordingly, these parameters for the $W\ 5p$ area are approximately found to be ($W\ 4f_{7/2}$ x 0.148) and the position is found to be $W\ 4f_{7/2}$ + 5.7 eV. It is important to note that the FWHM of this W 5p component could not be as simply restricted.

The Se 3d region is measured between ~51–61 eV. Similar to the MoSe$_x$ discussed for MoSe₂, the WSe₂ $Se\ 3d_{5/2}$ is fitted at 54.8 eV with the WSe$_x$ component at 55.3 eV. For WSe₂, as with all the selenide TMDs, there is a higher potential error in the specific binding energy assignment of each peak due to the overlap of the C 1s and a Se Auger component.

*Raman*

The Raman spectra for WSe$_2$ are shown in Fig.6 (c-e). WSe$_2$ has the 2H structure and the spectrum is dominated by the $E^1_{2g}$ and $A_{1g}$ Raman modes. For WSe$_2$ these are at very similar frequencies resulting in an intense convoluted peak.[66]

Raman of WSe2 is well studied under multiple excitation wavelengths in literature.[94, 95] The 532 nm excitation Raman spectrum of WSe$_2$ has a peak for the combined $E^1_{2g}$ and $A_{1g}$ modes at ~254 rel cm$^{-1}$. This large complex peak also contains contributions from second-order modes, a major contribution from the $2LA(M)$ centred at ~260 rel cm$^{-1}$ which shows strong layer dependence,[94] and another second-order peak at ~242 cm$^{-1}$. There is a general lack of discussion of the broad features at ~130 rel cm$^{-1}$ in the current literature, this complex convoluted structure of multiple features is difficult to assign. The remaining features are the results of a variety of second-order components.

Raman excitation at 405 nm of WSe$_2$ yields a simpler spectrum with a reduced number of visible components possibly due to this excitation being further from resonance. The contribution from the $2LA(M)$ is greatly reduced in this spectrum resulting in a far narrower main peak at ~254 rel cm$^{-1}$. Conversely, the 633 nm Raman spectrum has a relatively larger contribution from the $2LA(M)$ resulting in a broader main Raman feature. The structure of the complex broad feature at ~130 rel cm$^{-1}$ is also notably different at this excitation wavelength.

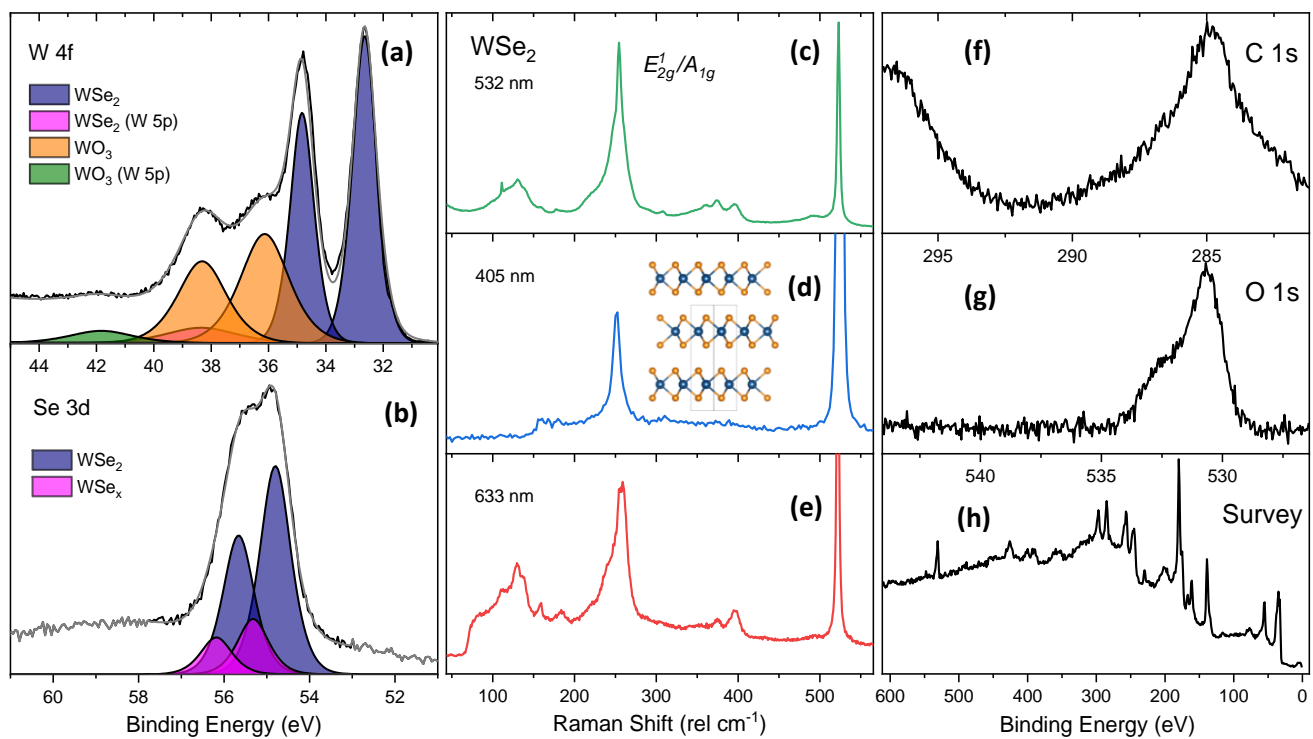

**Fig. 6** Peak-fitted XPS spectra of the W 4f **(a)** and Se 3d **(b)** core-levels. Raman spectra of WSe$_2$ acquired with 532 nm **(c)** 405 nm **(d)** and 633 nm **(e)** excitation energies with insert of the WSe$_2$ structure. XPS spectra of the C 1s **(f)**, O 1s **(g)** and survey **(h)** regions.

## PtSe$_2$

PtSe$_2$ is particularly interesting as it has only recently been studied for its 2D and thin film properties.[33, 96, 97] Given its recent emergence as a material of interest, there is a sparse but increasing volume of literature with high-quality characterisation of this material.

### *XPS*

The XPS regions for PtSe$_2$ characterisation are the Pt 4f (~74 eV) and the Se 3d (~33 eV) and their resultant data is illustrated in Fig. 7.

The energy region for measuring the PtSe$_2$ Pt 4f is ~69–82 eV. This region is fitted here with three Pt 4f doublets, as seen in Fig.7(a). The first doublet of $Pt\,4f_{7/2}$ is at 72.2 eV, attributed to a sub-stoichiometric PtSe$_x$ state. The presence of a somewhat correlated PtSe$_x$ component in the Se 3d spectrum increases confidence in this assignment. The symmetric nature of this peak indicates that it is not Pt metal, though this could potentially be an unidentified $PtO_x$ component.

The $PtSe_2$ $Pt\ 4f_{7/2}$ is fitted at 73.7 eV with 3.33 eV splitting. The third component is attributed to $PtO_2$ with $Pt\ 4f_{7/2}$ at 75.2 eV.

In an attempt to verify that this $PtO_2$ chemical state is in fact a Pt oxide, a Pt metal film was exposed to an oxygen plasma in a barrel asher. A comparison of the Pt 4f spectrum for $PtSe_2$, the oxygen plasma treated Pt film, and a pristine Pt metal film is provided in Fig. 8. It can be seen that after oxygen plasma treatment the film has two clear new chemical states, one with $Pt\ 4f_{7/2}$ at ~73 eV and the other at ~75 eV, these are attributed to a Pt(II), likely PtO, and also to a Pt(IV) attributed to $PtO_2$.[74] Similar oxide states are seen in many of the Pt TMDs measured here. This indicates that the $Pt\ 4f_{7/2}$ component at 75.2 eV for $PtSe_2$ is correctly assigned to a $PtO_2$ state.

The Se 3d core-level region is measured here between ~51–63 eV. The Se 3d region for $PtSe_2$ is the most complex of the Se 3d regions examined here. This core-level is fitted with seven components corresponding to four chemical states.

A major complication in this material is the coincidental overlap of the $Pt\ 5p_{3/2}$ with the low binding energy side of this spectrum, fitted here at approximately 53.7 eV. This broad feature is inconsistently apparent across the literature and is deserving of heightened scrutiny to determine the precise position and parameters of the $Pt\ 5p_{3/2}$ for $PtSe_2$. The $Pt\ 5p_{1/2}$ is not included here as the spin orbit splitting of the Pt 5p is 24 eV.

The $Se\ 3d_{5/2}$ for $PtSe_2$ is fitted at 55 eV. The $Se\ 3d_{5/2}$ corresponding to the previously discussed $PtSe_x$ state is fitted at 55.8 eV. The broad feature at ~59 eV is tentatively assigned here to a low-level of an undetermined Se oxide at 58.3 eV, there is potential for this broad feature to be associated to some extent with the $Pt\ 5p_{3/2}$.

*Raman*

$PtSe_2$ has the octahedral 1T structure shared by all the Pt TMDs measured here, and it can be assigned the $D_{3d}$ point group. The 532 nm excitation Raman spectrum of $PtSe_2$ matches very closely with the literature and shows the two prominent modes at 178 and 209 rel cm$^{-1}$, these are the $E_g$ and the $A_{1g}$ respectively. The shoulder at ~233 rel cm$^{-1}$ is from a combination of the

$A_{2u}$ and $E_u$ longitudinal optical modes.[97] These modes are sensitive to the film thickness with sharper peaks for lower thicknesses and a near absence of this mode for bulk samples.[97]

While the 405 nm excitation Raman spectrum of PtSe$_2$ maintains the two intense Raman modes at similar frequencies to the 532 nm, the ratio of these peaks is significantly different. The $A_{1g}$ mode is now slightly more intense than the $E_g$. Due to a low signal-to-noise ratio, it is difficult to ascertain if the broad feature at ~233 rel cm$^{-1}$ is present in this spectrum also. An apparent feature at ~156 rel cm$^{-1}$ appears for this wavelength, this is not a Raman feature and is simply a result of the spectrometer cut-off and can be seen at this position for all Raman spectra gathered with 405 nm excitation here.

The Raman spectra taken with 633 nm excitation is mostly similar to the 532 nm spectrum with the exception of the peak intensity ratio between the $E_g$ and $A_{1g}$ modes which shows a lower relative contribution from the $A_{1g}$ mode.

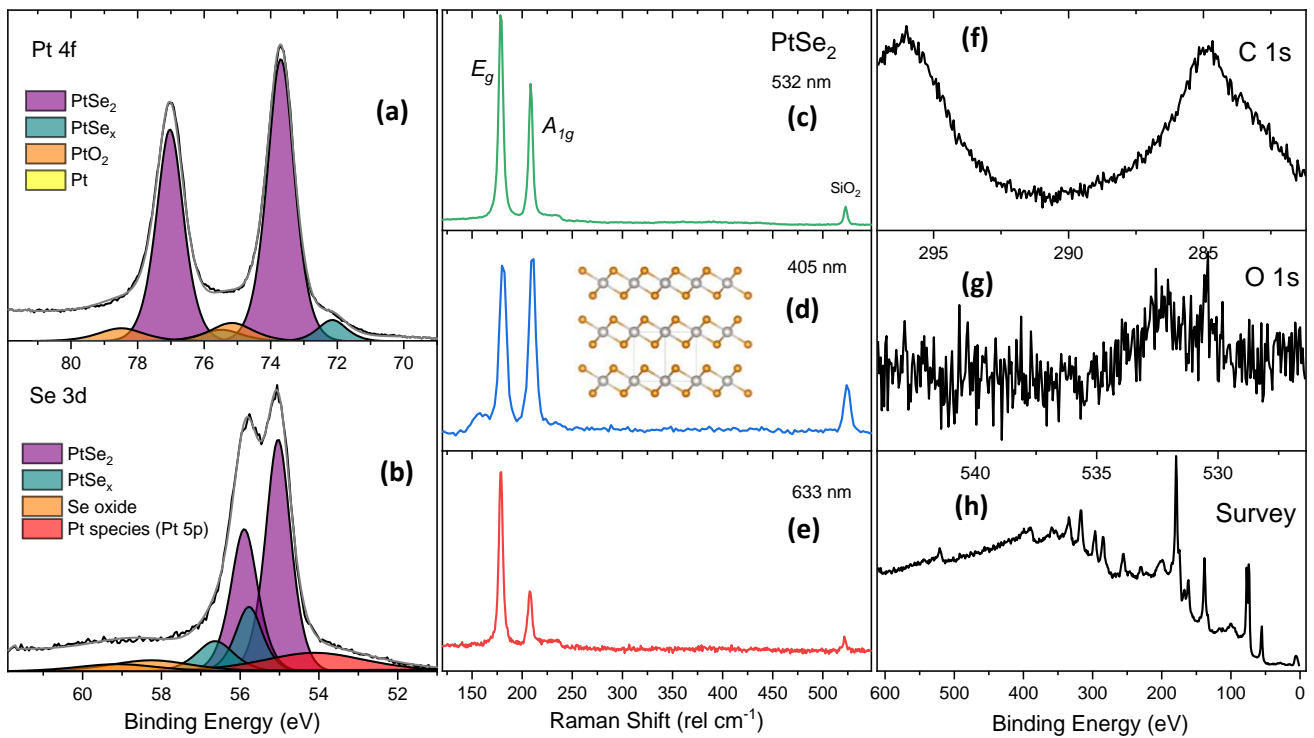

**Fig. 7** Peak-fitted XPS spectra of the Pt 4f **(a)** and Se 3d **(b)** core-levels. Raman spectra of PtSe$_2$ acquired with 532 nm **(c)** 405 nm **(d)** and 633 nm **(e)** excitation energies with insert of the PtSe$_2$ structure. XPS spectra of the C 1s **(f)**, O 1s **(g)** and survey **(h)** regions.

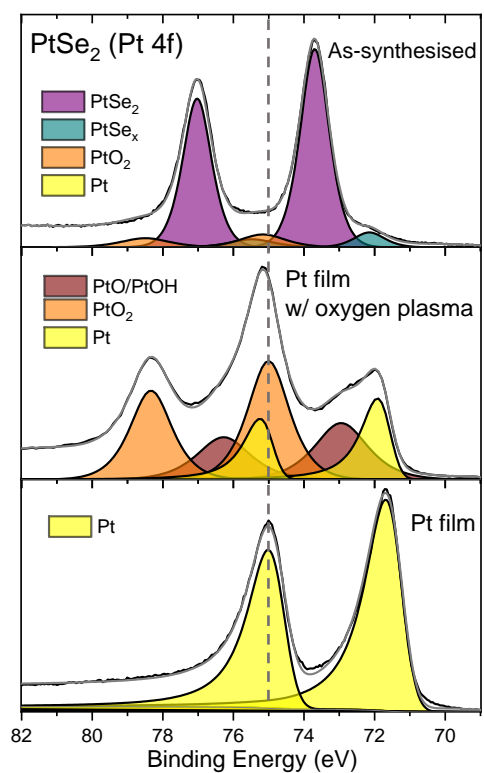

**Fig 8** A comparison of three Pt 4f XPS spectra. Showing Pt metal, plasma oxidised Pt metal, and PtSe$_2$ spectra.

**Telluride TMDs**

**MoTe$_2$**

Telluride TMDs are, as of the time of writing this study, by far the least studied of the TMDs. This may partly be due to the increased complexity of their synthesis and associated hazards due to the toxicity of Te.[98]

*XPS*

The XPS regions for MoTe$_2$ characterisation are the Mo 3d (~230 eV) and the Te 3d (~573 eV) and their resultant data is illustrated in Fig. 9.

Fig. 9(a) shows the 1T' MoTe$_2$ Mo 3d core-level spectrum, this is measured between 224–240 eV. This region is fitted with two doublets from two chemical states, MoTe$_2$ is the only Mo

TMD to not have overlapping XPS core-levels from the chalcogen. The MoTe$_2$ $Mo\ 3d_{5/2}$ component is fitted at 228.2 eV. Importantly, this doublet is asymmetric. Through a trial-and-error process this shape is found to be best fitted here by an asymmetric LA(1.1, 2.3, 2) line shape. This is indicative of metallic character of the film examined here and has been observed previously in the literature for 1T' MoTe$_2$, while 2H MoTe$_2$ has a symmetrical line-shape.[99]

The second doublet in the spectrum corresponds to the commonly observed MoO$_3$ and the $Mo\ 3d_{5/2}$ is fitted at 232.7 eV.

The Te 3d core-level region is measured between ~570–593 eV. The high spin-orbit splitting of this core-level of 10.38 eV necessitates this wide energy window. This region is fitted with four doublets, these will be described from low to high binding energy. The $Te\ 3d_{5/2}$ for MoTe$_2$ is fitted at 572.5 eV, the significantly higher FWHM of this compared to the MoTe$_2$ $Mo\ 3d_{5/2}$ component is due to the asymmetric peak shape of the Mo 3d component and the large difference in their respective binding energies.

An elemental Te component is fitted with $Te\ 3d_{5/2}$ at 573.2 eV. A sub-stoichiometric Te oxide, called TeO$_x$ here, is also fitted at 574.1 eV. The full rationale for the fitting and attribution of this component requires comparison across the three telluride TMDs measured here, as well as with the spectral data from the different anneal temperatures. The most common Te oxide TeO$_2$ has $Te\ 3d_{5/2}$ fitted here at 576.3 eV.

*Raman*

The Raman spectra of the distorted octahedral/monoclinic 1T' MoTe$_2$ film are shown in Fig.9 (c)-(e). A complication of the spectra for MoTe$_2$ and the other tellurides is the relatively low Raman shift of their characteristic Raman modes (~50–300 rel cm$^{-1}$) This can make accurate characterisation partially dependant on the low-frequency range of the spectrometer. The 532 and 633 nm wavelengths here are able to characterise all of the peaks in this range, the 405 nm had a higher spectrometer cut-off of ~140 rel cm$^{-1}$, therefore limiting the number of resolvable peaks.

1T' MoTe$_2$ has the low symmetry point group $C_{2h}^2$ giving it numerous low-frequency modes. The high level of complexity to the Raman analysis has resulted in telluride TMDs being a less settled science and there is still some ambiguity as to the assignment of modes seen in the Raman spectra.[100-104] The work by Song et al.[105] will serve as the basis for the majority of this limited Raman study.

Seven Raman peaks are measured in the 532 nm Raman spectrum in Fig. 9 (c), five of these at ~ 80, 89, 129, 164, and 258 rel cm$^{-1}$ are assigned the $A_g$ symmetry, the remaining peaks at ~ 111 and 192 rel cm$^{-1}$ are assigned the $B_g$ symmetry.

The impact of the Rayleigh scattering filter on the 405 nm spectrum cuts off the peaks below the main $A_g$ peak, leaving only the 164 and 258 rel cm$^{-1}$ peaks for the $A_g$ symmetry. The 192 rel cm$^{-1}$ $B_g$ feature is maintained with this excitation wavelength and appears to show a higher relative intensity than for 532 nm.

Moreover, the 633 nm spectrum shows a similar spectrum to the 532 nm with several exceptions. The 192 rel cm$^{-1}$ $B_g$ Raman peak is no longer detected. The Raman modes between 85–150 rel cm$^{-1}$ are all significantly broader with further overlap of the peaks making accurate assignment more difficult.

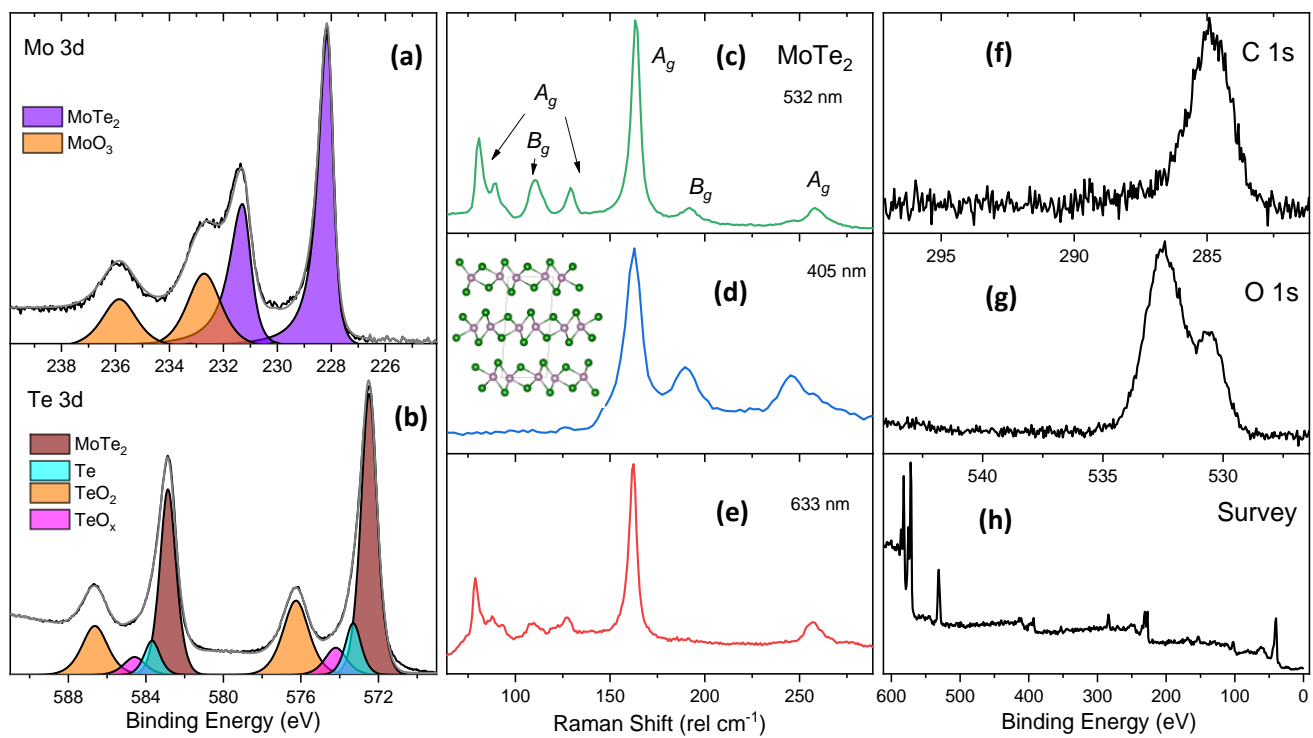

**Fig. 9** Peak-fitted XPS spectra of the Mo 3d **(a)** and Te 3d **(b)** core-levels. Raman spectra of MoTe$_2$ acquired with 532 nm **(c)** 405 nm **(d)** and 633 nm **(e)** excitation energies with insert of the MoTe$_2$ structure. XPS spectra of the C 1s **(f)**, O 1s **(g)** and survey **(h)** region

**WTe₂**

*XPS*

The main characteristic XPS regions for WTe$_2$ are the W 4f (~33 eV) and Te 3d (~573 eV). Regarding XPS characterisation, WTe$_2$ is the most complicated of the TMDs discussed in this work due to the overlap of the W 4f, W 5p, and the Te 4d core-level regions. Thus, it is found that measuring between ~30–48 eV allowed for the most appropriate fitting here. This region, as shown in Fig.10, is fitted with fourteen components from just five different chemical states.

The WTe$_2$ $W\ 4f_{7/2}$ is fitted at 31.7 eV, with an asymmetric peak shape. Through trial and error this is most accurately described here by a LA(1.3, 2.5, 11) line shape. The corresponding $W\ 5p$ component used the fixed parameters described for WSe$_2$ and is fitted at 37.4 eV.

There is a prominent oxide contribution in the as-synthesised material in the form of WO$_3$ with its $W\ 4f_{7/2}$ fitted at 35.8 eV and its $W\ 5p_{3/2}$ at 41.5 eV.

Turning to the Te 4d components, these are subject to significant overlapping. Accurate fitting of these components required comparative analysis with the commonly studied Te 3d region. Four sets of doublets are fitted for the Te 4d with spin-orbit splitting of 1.47 eV. The WTe$_2$ $Te\ 4d_{5/2}$ is fitted at 40.4 eV. The second doublet is attributed to an elemental Te component with $Te\ 4d_{5/2}$ at 40.87 eV. The most common oxide of Te, TeO$_2$, is significantly shifted from the other peaks and is fitted with $Te\ 4d_{5/2}$ at 44.2 eV. A low-intensity component is found with $Te\ 4d_{5/2}$ at 42.2 eV, this is attributed to a sub-stoichiometric oxide of Te, called TeO$_x$ here.

These components are all represented in the Te 3d core-level region in Fig.10 (b), measured between ~570–593 eV. The WTe$_2$ $Te\ 3d_{5/2}$ is fitted at 572.4 eV, with the elemental Te and TeO$_x$ peaks at 573 and 574.1 eV respectively. The TeO$_2$ component is well separated and is fitted at 576.2 eV.

*Raman*

The Raman spectra of WTe$_2$ at the three excitation wavelengths are given in Fig.10 (c–e). Distorted octahedral/orthorhombic T$_d$ WTe$_2$ has the $C_{2v}^7$ point group, it has 12 atoms in the unit cell giving rise to several Raman modes. Similar to MoTe$_2$, the Raman spectrum of WTe$_2$ is quite complex and has not been as extensively studied as other 2D TMDs. The assignment of the Raman peaks here is based on several recent works.[46, 106, 107]

The 532 nm excitation spectrum has seven clearly discernible component peaks. Five of these are related to an $A_1$ symmetry, with peaks at ~ 82, 116, 134, 163, and 211 rel cm$^{-1}$. The other two modes are from an $A_2$ symmetry and are at ~ 91, and 112 rel cm$^{-1}$.

The 405 nm spectrum is of limited utility as the modes below 150 rel cm$^{-1}$ are cut off. The low signal-to-noise ratio leaves the $A_1$ peak at 163 rel cm$^{-1}$ and the $A_2$ peak at 211 rel cm$^{-1}$ as only partly discernible.

Using 633 nm excitation spectra here also suffered from a low signal-noise ratio. Accounting for this, little substantial difference from the 532 nm spectrum can be distinguished, besides a relative reduction in the intensity of the $A_2$ components.

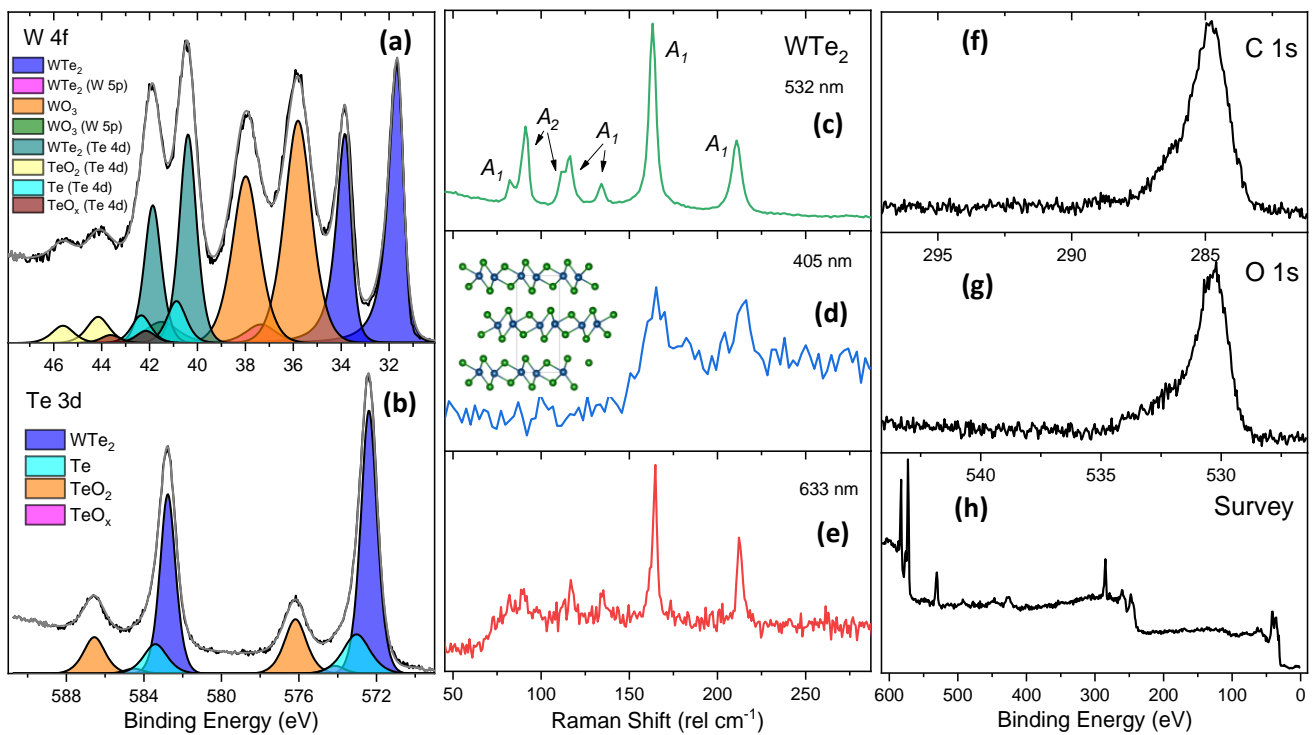

**Fig. 10** Peak-fitted XPS spectra of the W 4f **(a)** and Te 3d **(b)** core-levels**.** Raman spectra of WTe$_2$ acquired with 532 nm **(c)** 405 nm **(d)** and 633 nm **(e)** excitation energies with insert of the WTe$_2$ structure. XPS spectra of the C 1s **(f)**, O 1s **(g)** and survey **(h)** regions.

**PtTe$_2$**

As was the case for the sulfides and selenides, PtTe$_2$ is the least studied of the telluride TMDs discussed herein. PtTe$_2$ is metallic in the bulk form and has been artificially synthesised since at least 1897.[108]

*XPS*

The Pt 4f (~73 eV) and the Te 3d (~573 eV) core-levels are used for characterisation of PtTe$_2$. The resultant data are illustrated in Fig. 11

The measured energy window for the Pt 4f here is ~68–82 eV. This region is fitted with six components from three chemical states. Unlike for the other Pt based TMDs, an asymmetric $Pt\ 4f_{7/2}$ peak is fitted at 71.3 eV and is attributed to Pt metal. It is not determined if this is the result of unconverted starting material or was a product of degradation of the material prior to the XPS measurements. The $Pt\ 4f_{7/2}$ corresponding to PtTe$_2$ is fitted at 72.7 eV. A broader low energy doublet, seen for PtS$_2$ and PtSe$_2$ also is assigned as PtO$_2$ and is fitted with the $Pt\ 4f_{7/2}$ at 73.8 eV.

The Te 3d core-level region is measured between ~569–591 eV. This region is measured to include eight components from four chemical states of Te. These are largely similar to those described for the other TMD tellurides discussed previously. The $Te\ 3d_{5/2}$ for PtTe$_2$ is fitted at 572.9 eV. The elemental Te and TeO$_x$ components are fitted at 573.4 and 573.9 eV respectively. The most common Te oxide, TeO$_2$ is fitted at 575.6 eV.

*Raman*

The Raman spectra of PtTe$_2$ are given in Fig. 11(c)-(e). Unlike the other tellurides, PtTe$_2$ is isostructural with its corresponding sulfide and selenide TMDs. It has a 1T structure giving the $D_{3d}$ point-group symmetry and corresponding Raman modes, and therefore a significantly simpler Raman spectrum than the other 2 tellurides.[109]

The 532 nm Raman spectrum shows two clear peaks, these are the $E_g$ mode at ~112 rel cm$^{-1}$ and the $A_{1g}$ mode at ~159 rel cm$^{-1}$. As was the case for the other telluride TMDs, the 405 nm Raman spectrum is of much lower utility as much of the region of interest is below the cut-off. Lastly, the 633 nm excitation Raman of PtTe$_2$ shows the same two Raman modes present for 532 nm excitation at very similar frequencies.

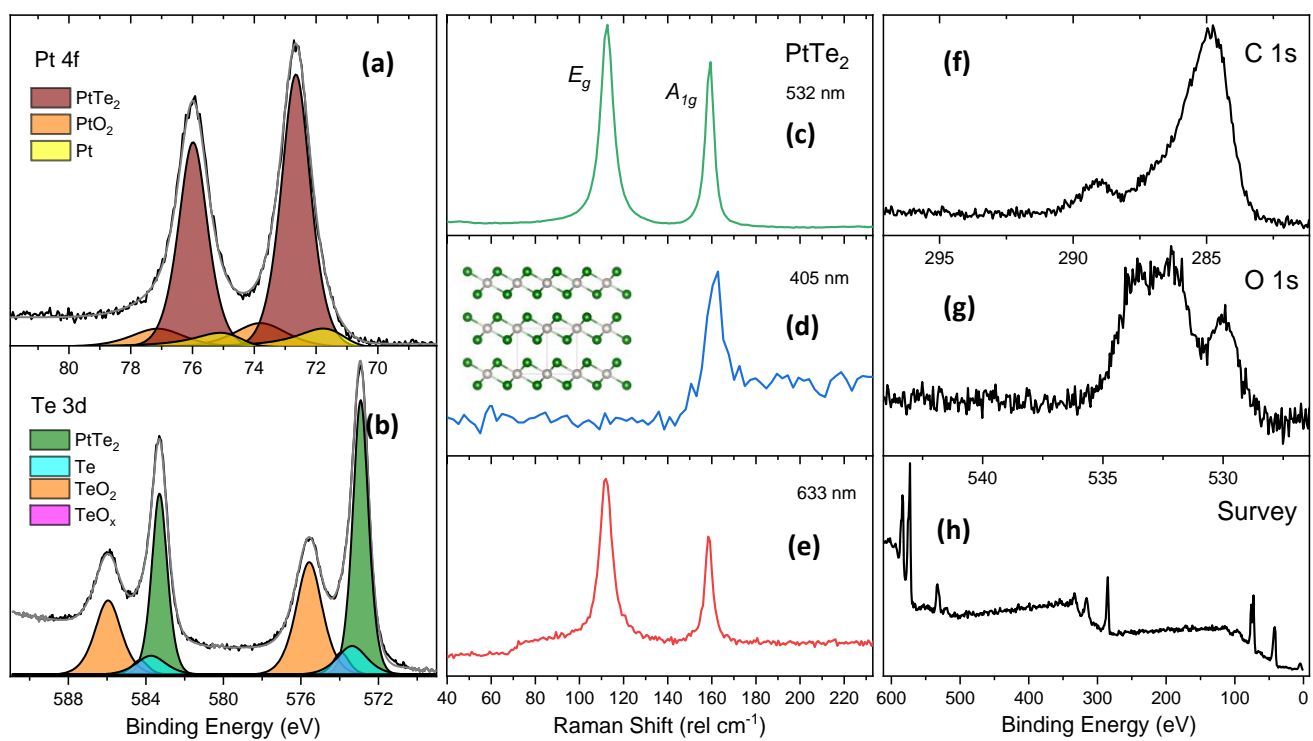

**Fig. 11** Peak-fitted XPS spectra of the Pt 4f **(a)** and Te 3d **(b)** core-levels. Raman spectra of PtTe$_2$ acquired with 532 nm **(c)** 405 nm **(d)** and 633 nm **(e)** excitation energies with insert of the PtTe$_2$ structure. XPS spectra of the C 1s **(f)**, O 1s **(g)** and survey **(h)** regions.

**Stoichiometry**

One of the most common uses of XPS in 2D materials literature is to measure the stoichiometry of the material. While XPS is one of the few methods through which this is possible, it is not a trivial process.[24] Quantitative analysis with XPS requires multiple factors to be considered including the transmission function, relative-sensitivity factors, sample cleanliness, sample thickness, etc. These are in combination with key fitting choices, which each can have significant impacts on any calculated stoichiometry. Due to these factors, accurate measurement of material stoichiometry is most applicable to experiments in ideal conditions with very high-quality material in pristine environments.

Intentionally the XPS analysis of 2D TMDs presented here uses materials which have been exposed to ambient and have surface contamination, defects, and are partially oxidised. This reflects more accurately many of the XPS spectra seen in the literature for these materials and common laboratory conditions.

Due to their all-surface nature, the properties of 2D materials can be greatly affected by variations in their stoichiometries - defects, vacancies and adatoms are all well known to be common occurrences.[110] While it can be tempting to draw significant conclusions from XPS-determined stoichiometry, the error in these calculations can often be greater than any real variation from ideal stoichiometry in the material. However, this does not mean that XPS-calculated stoichiometry is not worthwhile or valuable in non-ideal conditions, semi-quantitative analysis of the stoichiometry serves a key role when validating the XPS peak fitting for a material.

The calculated stoichiometry from the XPS of the nine TMD materials discussed thus far are presented in Fig.12. An average stoichiometry of 1.92 with standard deviation of 0.13 gives an indication of the typical error in these calculations. When estimating error, the $PtS_2$ calculated stoichiometry of 1.69 is the only material clearly indicating a definitive chalcogen deficiency. For $PtS_2$ this is likely a combination of the presence of PtS in the material, the complex S 2p XPS core-level, and the multiple overlapping components leading to inherently higher potential error in analysis.

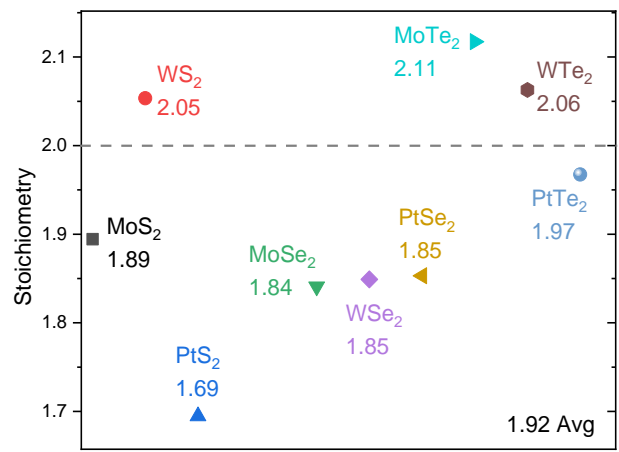

**Fig. 12** Calculated stoichiometry values for the as-synthesised TMD films

**Thermal Stability and Characterisation of TMDs**

An effective method to aid in the detection and assignment of the various features of an XPS spectrum is to analyse a series of data, this can help to differentiate between components and indicate their origin e.g. oxide or TMD.

In this experiment part of each as-synthesised film was annealed in ambient conditions at temperatures up to 400 °C, to gradually oxidise or otherwise decompose the material. These annealed films were then characterised with XPS and Raman spectroscopy. Optical images of the sample surfaces are presented in Fig.S4. This method not only improves the reliability of the fitted XPS spectra of all the thin-film materials measured here but also provides a better understanding of the oxidation tendencies of the materials. The in-air stability of each material here is only indicative due to the large increments between each anneal temperature.

The large volume of data generated in this section provides the greatest insights when viewed comparatively, and so the XPS and Raman data for each set of TMDs, categorized by chalcogen, will be displayed together (e.g. comparison of $MoS_2$, $WS_2$ and $PtS_2$). New insights can also be gained by comparing these data by transition metal (e.g. comparison of $MoS_2$, $MoSe_2$ and $MoTe_2$), this will be shown and discussed subsequently.

Due to the large number of figures in this study, two data presentation decisions are made. Firstly, the XPS spectra of the C 1s, O 1s, and survey region for all the materials are included in the supplementary information, Fig. S5. Secondly, the assembled Raman data is displayed closely together for each material to aid in direct comparison. This can hinder the readability of the finer details in some spectra. Consequently, all the Raman spectra are also displayed individually in Fig. S7–10, grouped by Raman excitation energy.

**Transition Metal Sulfides**

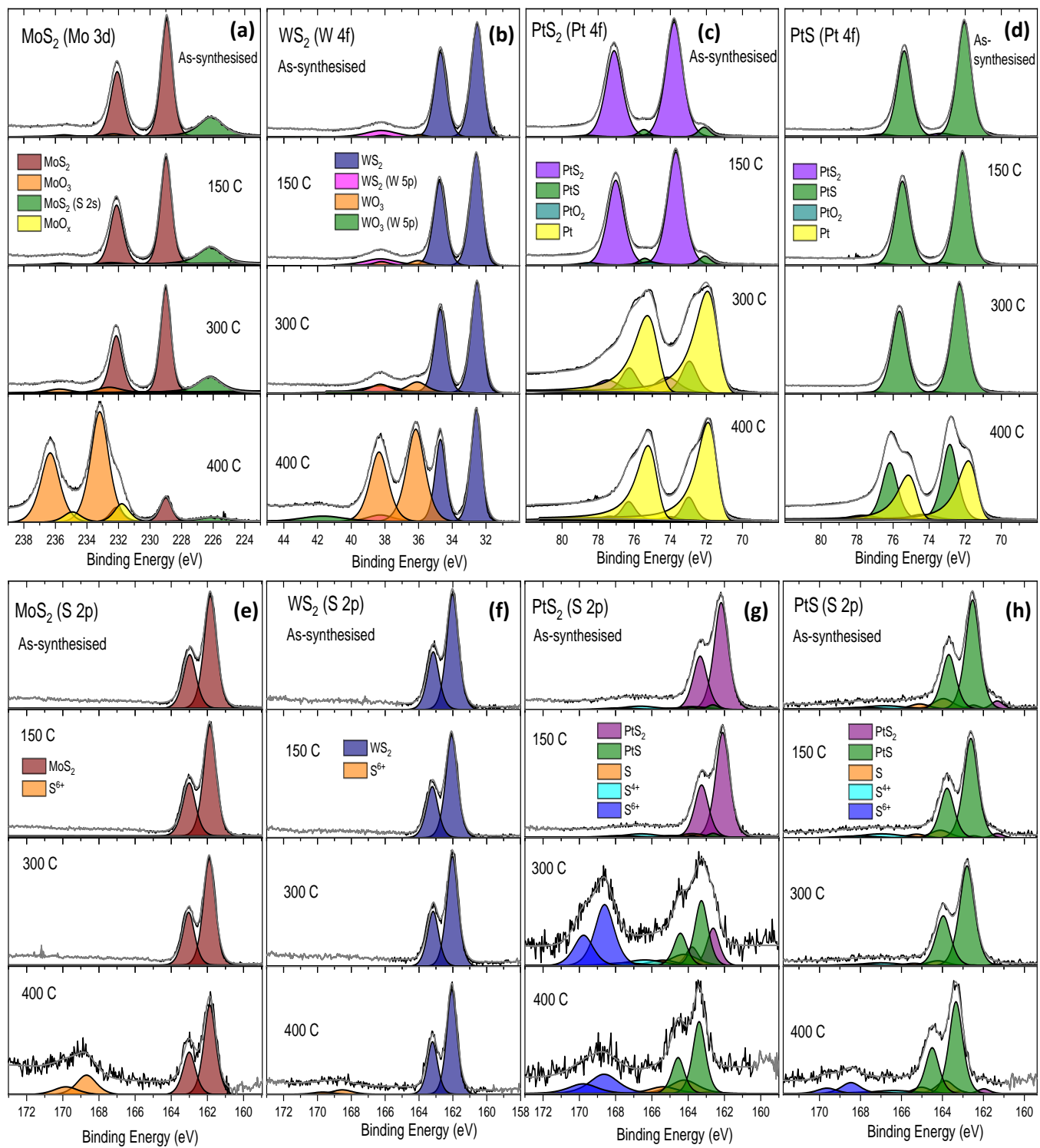

**Fig. 13** XPS spectra for the sulfide materials showing multiple annealing steps. Transition metal core-levels **(a)-(d)** and S 2p core-levels **(e)-(h)** from each material.

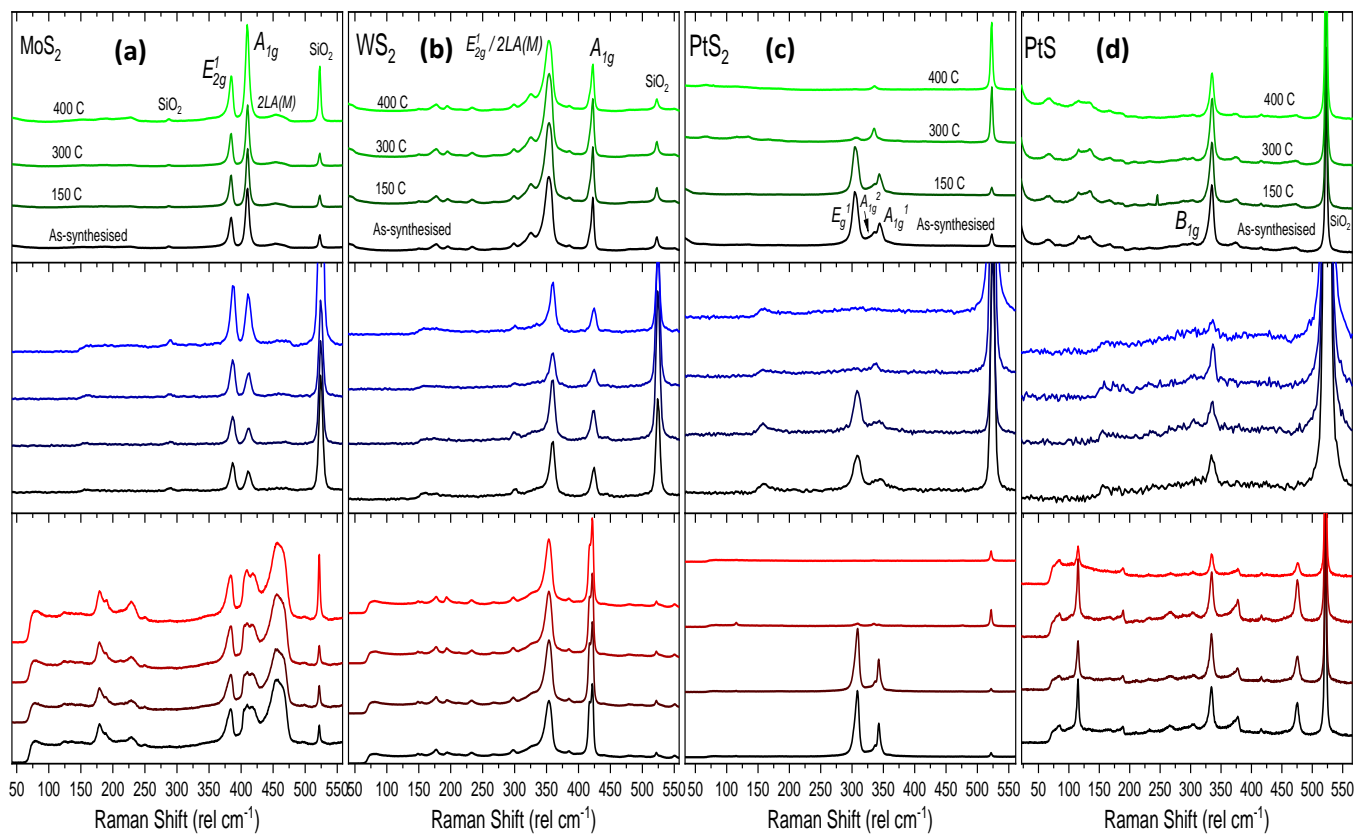

**Fig. 14** Raman spectra for the sulfide materials after each anneal acquired with three laser excitation energies, green for 532 nm, blue for 405 nm, and red for 633 nm. $MoS_2$ **(a)** $WS_2$ **(b)** $PtS_2$ **(c)** PtS **(d)**

The XPS spectra for the previously discussed $MoS_2$ Mo 3d and S 2p core-levels are presented in Fig.13 (a, e) respectively. Both regions show significant stability up to 300 °C where there is an increase in $MoO_3$ contribution to ~10%. This indicates that we have passed the threshold for oxidation, aligning with reports in literature of this occurring at ~290 °C in ambient.[111] There are no signs of oxidation in the S 2p core-level region up to 300 °C, indicating that the sulfur lost from the oxidation of $MoS_2$ is being lost to the ambient environment during the anneal or alternatively being lost to the vacuum during pump-down prior to the XPS measurement.

Both spectra following annealing at 400 °C show pronounced oxidation with only 10% of the surface molybdenum atoms being in the $MoS_2$ state. The majority of the oxidation is through the formation of $MoO_3$. A secondary oxide labelled as $MoO_x$ forms at some stage between 300–400 °C, this is fitted at ~232 eV, this oxide has been previously observed but a conclusive assignment is difficult.[112]

The S 2p region develops an oxidation-induced component in the 400 °C spectra at ~168.7 eV, attributed to $S^{6+}$, likely in a sulfate molecule. This oxide component only accounts for ~30% of the surface sulfur, with the remaining 70% being $MoS_2$. This indicates that sulfur oxide formation is a secondary process in $MoS_2$ degradation.

The Raman spectra of $MoS_2$ at three excitation wavelengths are shown in Fig.14(a). For the 532 nm spectrum it is noteworthy that the $MoS_2$ Raman is almost imperceivably changed with increasing anneal temperature with no oxide peaks present. The starkest change in the spectrum is seen after the 400 °C anneal, but only in the changing relative intensity of the $SiO_2$ Raman peak at ~520 rel cm$^{-1}$ when compared to the $MoS_2$ Raman modes.

While the XPS data makes it clear that the surface of the $MoS_2$ film is largely oxidised after the 400 °C anneal, the Raman data tells us that there is still a proportion of the initial $MoS_2$ which has been undisturbed. The difference between penetration depth for Raman and XPS allows us to gain insight of both the surface and substrate facing sides of the $MoS_2$ film, with XPS measuring the first few nanometres of the surface while Raman is generally probing through the entire thin film of TMD.

The results of ambient thermal oxidation of $WS_2$ are similar to those of $MoS_2$. XPS of the W 4f core-level for $WS_2$ films, Fig.13(b), shows very little oxide formation on the surface up to 300 °C, with only ~12% of W atoms in the $WO_3$ chemical state. At 400 °C the $WS_2$ film now has 60% $WO_3$ surface oxide, this makes $WO_3$ the least oxidised of any of the TMDs measured here after the 400 °C anneal. Similarly, the S 2p shows no major changes until 400 °C where a second component at ~168.5 eV appears. This low-intensity component only accounts for ~8% of the remaining sulfurs, the majority of sulfur atoms are therefore lost from the sample rather than forming surface oxides at this temperature.

The Raman spectra of $WS_2$ show no distinctive changes after annealing at any temperature for the three excitation wavelengths used here. The relative intensity of the $SiO_2$ signal is mostly unchanged, indicating there is not substantial thinning or broad lowering in film quality, this further indicates the remarkable stability of the $WS_2$ film in this high temperature ambient environment.

An important impact of this data is the disconnect of the Raman and XPS data for both $WS_2$ and $MoS_2$. Many reports utilising these materials, which are not characterisation-focused reports, solely rely on one characterisation technique to determine TMD nature and quality, primarily Raman spectroscopy. This is clearly unsatisfactory for thin films where a minimum of two complementary characterisation techniques are necessary before any determinations of quality can be realistically made. In certain circumstances it may be mostly satisfactory for characterisation of monolayer films, but even in that case a single characterisation method may result in erroneous conclusions.

The results for $PtS_2$ and PtS show that these are notably more reactive materials than $MoS_2$ or $WS_2$. After annealing at 150 °C there is no change for either $PtS_2$ or PtS, in either the Raman or XPS. PtS is stable after the 300 °C anneal with the only change being a reduction in the amount of impurities, including $PtS_2$, detected in the S 2p XPS region. $PtS_2$ shows substantial instability after this 300 °C annealing step with the composition of the film changing completely. The Pt 4f region shows that now ~75% of the surface Pt atoms have been decomposed into elemental Pt. Interestingly, the second largest contribution in this film is now attributed to a PtS state, with only ~10% of $PtS_2$ remaining.

The S 2p region also shows this change with a large PtS state accompanied by a shoulder from the $PtS_2$. Contrasting with the results for the other sulfide films, there is a large $S^{6+}$ component formed here. The Raman spectrum of the sample is transformed after the 300 °C anneal with the single peak of PtS at ~335 rel cm$^{-1}$ now the most prominent feature of the spectrum. This combination of data indicates that Pt oxides are not favourable in this scenario of ambient thermal oxidation, formation of Pt metal and reduction to PtS are the primary degradation pathways of $PtS_2$.

The sample annealed at 400 °C shows a continuation of this process for the $PtS_2$ film with all the remaining $PtS_2$ being converted to PtS or Pt metal, the $S^{6+}$ contribution is lower than for 300 °C. The Raman signal for this film is now very low in intensity as this high temperature results in some degradation of even the relatively stable PtS formed from the initial $PtS_2$ film. The post-400 °C Raman spectra for $PtS_2$ and PtS are included in Fig.S7-10, the striking similarity of these Raman spectra is apparent and reinforces the hypothesised process of $PtS_2$ converting to PtS with annealing. The XPS spectra for PtS after the 400°C anneal firmly establish it as being one of the more thermally stable materials investigated, with an approximate 50/50 split of the Pt region being PtS and Pt metal. The S 2p region changes with oxidation to include a minor component of $S^{6+}$ similar to $PtS_2$.

**Transition Metal Selenides**

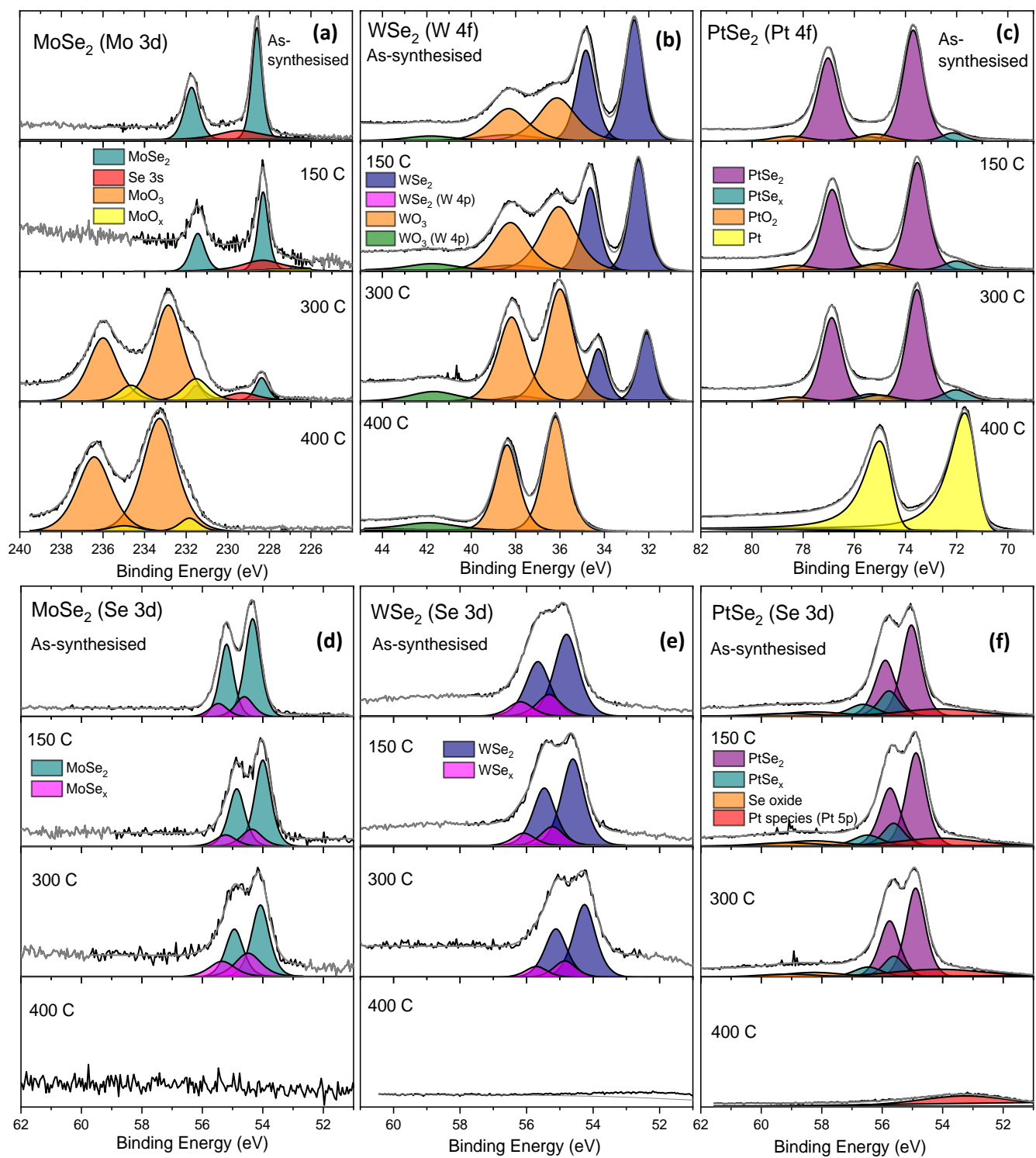

**Fig. 15** XPS spectra for the selenide TMDs showing multiple annealing steps. Transition metal core-levels **(a)-(c)** and Se 3d core-levels **(d)-(f)** from each material.

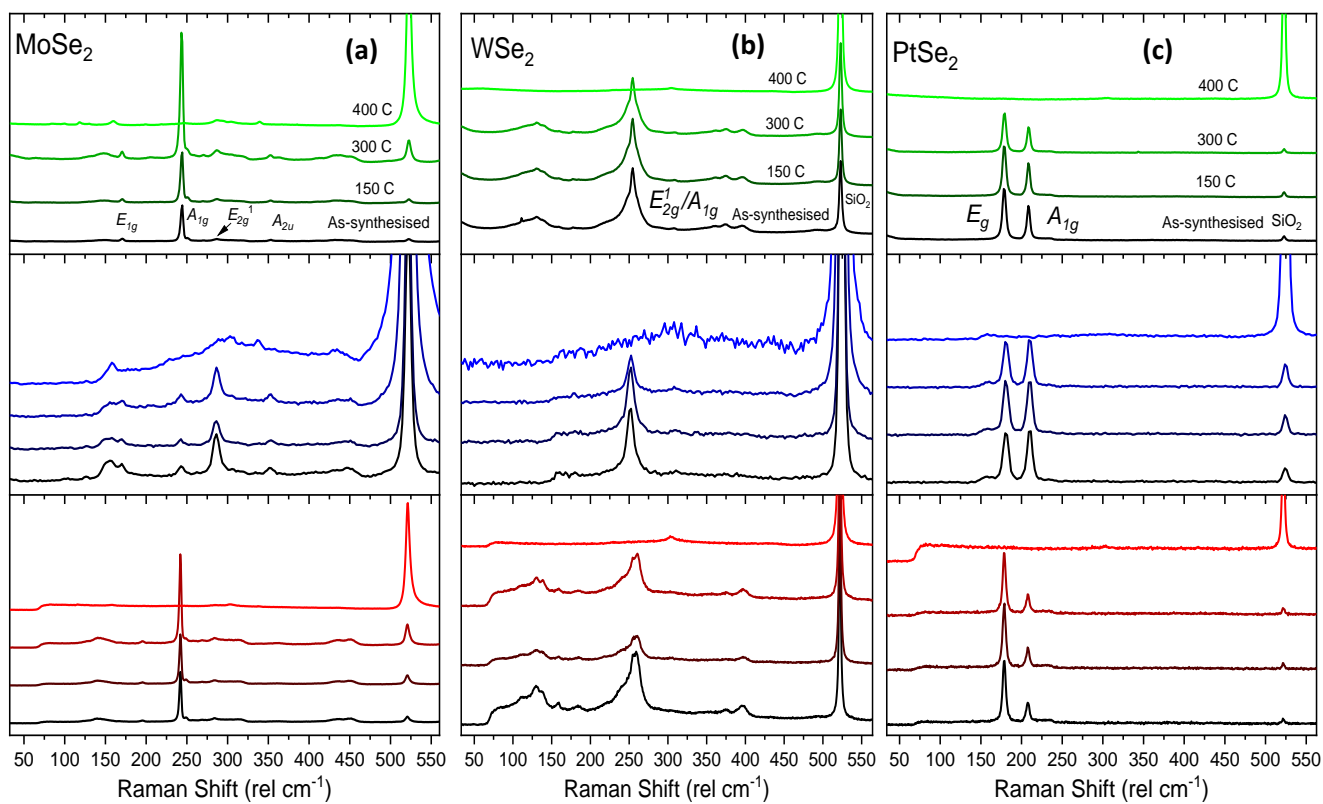

**Fig. 16** Raman spectra for the selenide TMDs after each anneal acquired with three laser excitation energies, green for 532 nm, blue for 405 nm, and red for 633 nm. MoSe$_2$ **(a)** WSe$_2$ **(b)** PtSe$_2$ **(c)**.

The thermal stability of the TMD selenides is compared in Fig.15 and Fig.16. The Mo 3d and Se 3d XPS core-levels for MoSe$_2$ are shown in Fig.5.19 **(a)-(d)**. MoSe$_2$ shows no formation of any oxides after 150 °C annealing in air indicating stability up to this point.

Annealing at 300 °C results in heavy oxidation, with only ~8% of the Mo atoms being in the form of MoSe$_2$. Two prominent oxides are formed, MoO$_3$ at ~233 eV and an unidentified oxide, likely the same one present in 400 °C annealed MoS$_2$, at ~232 eV. The corresponding Se 3d spectrum shows no sign of any new chemical states but shows broadening of the XPS peaks by ~15% when compared with the pristine material indicating a growing bond diversity and a lowering of crystallinity as Se leaves the TMD.

The results after 400 °C annealing reveal that MoSe$_2$ is not stable at this temperature as it was undetectable. The Mo 3d region consists of only the two previously-discussed oxides, with the MoO$_3$ increasing in relative amount, indicating its greater stability than the unidentified oxide. The Se 3d region shows no Se atoms remaining.

Raman analysis of these films confirms the loss of MoSe$_2$. There are only minor changes in the spectra for all excitation wavelengths up to the 300 °C anneal. Consistent with the XPS data, there is a loss of MoSe$_2$ Raman signal at 400 °C. Several new Raman peaks appear at this step, the most prominent peaks are at ~119, 160, 287, 304, and 339 rel cm$^{-1}$. These peaks are known to correspond to orthorhombic MoO$_3$, confirming the XPS conclusions.[113] This draws a sharp contrast between the sulfides and the selenides wherein the 400 °C anneal causes oxidation of the surface layers of sulfide TMDs, but selenide films are completely oxidised at this temperature, showing the general higher stability of the sulfides.

The pristine WSe$_2$ sample measured here is found to be substantially oxidised on the surface before annealing. This starting oxide complicates any definitive statements on material stability but nonetheless can still serve to portray the most likely degradation pathway with ambient annealing. XPS analysis of WSe$_2$ shows a clear trend with a gradually decreasing amount of WSe$_2$ with a corresponding increase in WO$_3$ up to 300 °C. A notable gradual shift to lower binding energy is observed in the Se 3d indicating a potential p-type doping as a result of oxidation.[114]

Similar to MoSe$_2$, there is no WSe$_2$ remaining after annealing at 400 °C. The W 4f after 400 °C only has signal pertaining to WO$_3$ remaining, the FWHM of this $W\ 4f_{7/2}$ narrows substantially across annealing steps with a change from 2 -> 1.86 -> 1.47 -> 1.22 respectively with increasing anneal temperature.

PtSe$_2$ shows the best thermal stability of any of the selenide materials studied here. The Pt 4f and Se 3d show marginal change for ambient anneal temperatures up to 300 °C, this indicates a relatively high stability. The limit to this is found to be between 300 and 400 °C as both XPS core-levels show the complete loss of PtSe$_2$ after annealing at 400 °C, this is consistent with the other selenides. Unlike the other materials, Pt TMDs do not form oxides during ambient annealing but Pt metal.

The Raman spectra of PtSe$_2$ are consistent with the XPS data and show that there are minimal changes to the spectra after 300 °C but a complete loss of PtSe$_2$ Raman signal after the 400 °C anneal.

# Transition Metal Tellurides

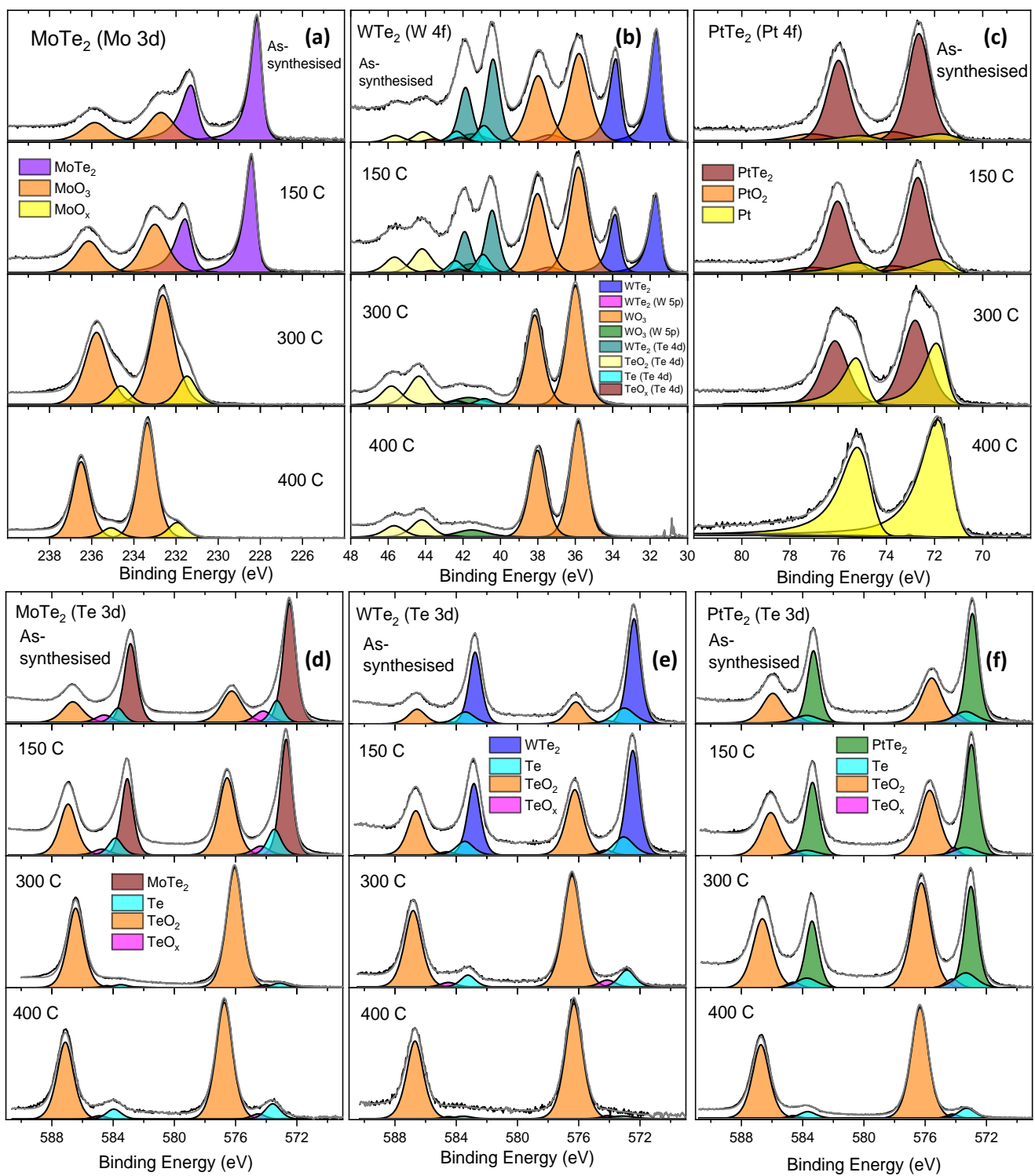

**Fig. 17** XPS spectra for the telluride TMDs showing multiple annealing steps. Transition metal core-levels **(a)-(c)** and Te 3d core-levels **(d)-(f)** from each material.

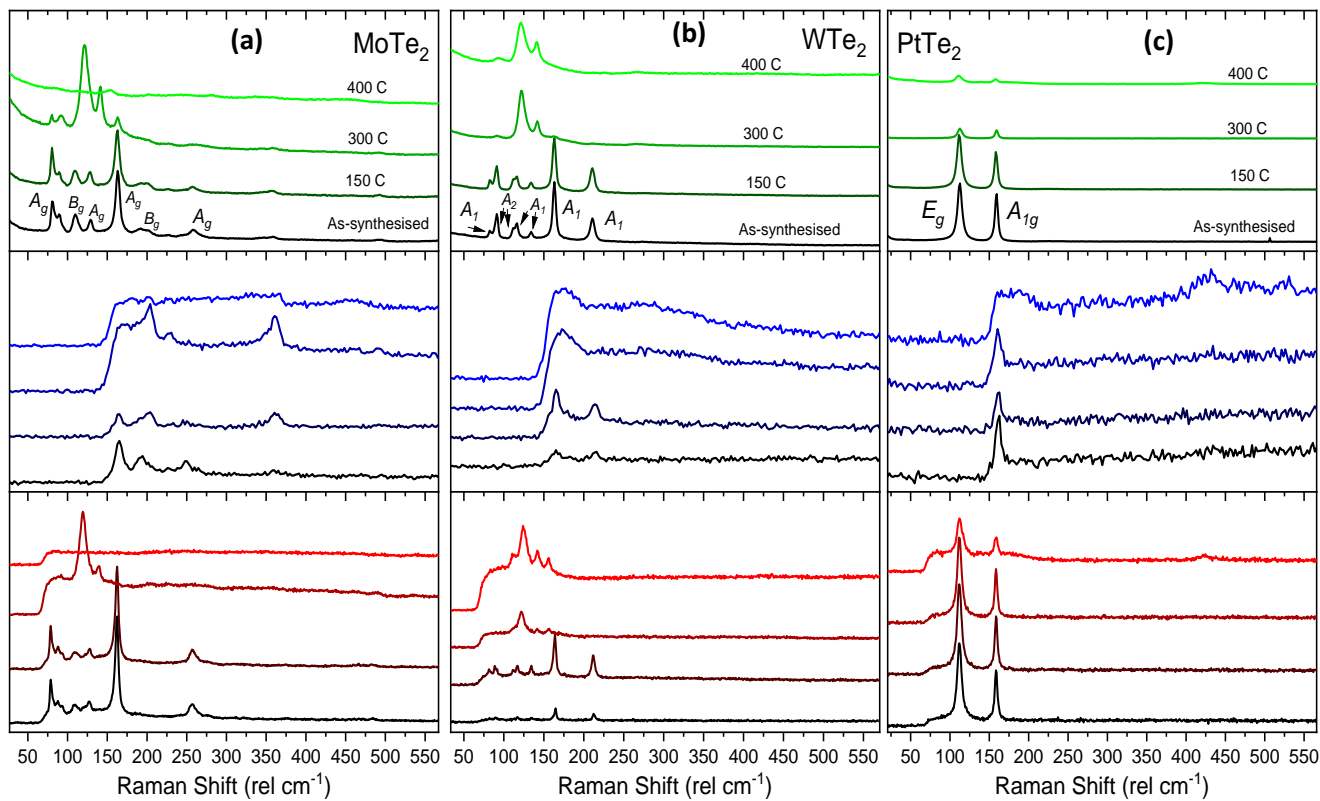

**Fig. 18** Raman spectra for the telluride TMDs after each anneal acquired with three laser excitation energies, green for 532 nm, blue for 405 nm, and red for 633 nm. MoTe$_2$ **(a)** WTe$_2$ **(b)** PtTe$_2$ **(c)**.

The XPS and Raman data for MoTe$_2$ films are given in Fig.17 **(a)-(d)** and Fig.18 **(a)-(c)**. For the MoTe$_2$ data a similar trend to MoSe$_2$ is observed for the Mo 3d core-level. The pristine film contained ~30% of the surface Mo atoms as MoO$_3$, there is a gradual increase with each anneal temperature due to its poor ambient thermal stability. Similar to the other Mo TMDs, MoO$_x$ is formed for the 300 and 400 °C anneal temperatures. Any traces of MoTe$_2$ XPS signal are absent after the 300 °C anneal in the Mo 3d region.

The Te 3d XPS core-level for MoTe$_2$ shows consistent results with a complete loss of MoTe$_2$ signal after the 300 °C anneal. The relative amount of TeO$_2$ increases for each anneal temperature. XPS signal from elemental Te and TeO$_x$ are found to persist across all anneals with consistently low relative amounts.

The Raman spectra from these films, shown in Fig.18 **(a)**, reveal an interesting picture and highlight substantial change across the annealing steps. As a result of the different synthesis procedure for the telluride TMD films, these films show greater surface roughness, which is further increased by the annealing experiments conducted here. This increased roughness can hamper the multiple spectra averaging process used during the Raman analysis, resulting in greater variability and noise in the telluride results than for the other TMD films. The 532 nm excitation Raman spectra for $MoTe_2$ show a loss of the majority of the $MoTe_2$ characteristic modes after the 300 °C anneal. New modes with pronounced peaks at ~121 and 141 rel cm$^{-1}$ appear and these are consistent with previous reports of decomposition of $MoTe_2$ films, with these peaks assigned to Te rather than $TeO_2$.[115-117]

Raman spectra acquired using 405 nm excitation show a substantially different spectra due to the spectrometer cut-off at this wavelength.

The 633 nm Raman spectra show largely similar results to the 532 nm, with a complete loss of $MoTe_2$ signal after the 300 °C anneal and the presence of the Raman peaks corresponding to Te.

Ambient thermal stability of $WTe_2$ shows appreciably similar results to $MoTe_2$, which firmly establishes the trend of Mo and W TMDs being chalcogen limited in their stability.

The W 4f core-level is the standard literature core-level when characterising W materials, this region also contains the W 5p adding to its complexity. The region is further complicated for the case of $WTe_2$ due to the coincidental overlap of the Te 4d core-level. This necessitates a wide energy window fitted with twelve components. Similar to $MoTe_2$, for the 150 °C anneal a reduction in the amount of TMD signal is noticed with an increase in oxide content, in this case $WO_3$. There is no remaining TMD signal after the 300 °C anneal, and the 400 °C anneal leaves $WO_3$ and $TeO_2$ as the only states detected on the surface.

The Te 3d core-level is the primary characteristic XPS region for tellurium compounds and is shown in Fig. 17 **(e)**. This shows a remarkably similar trend to the Te 3d of $MoTe_2$ except for the 400 °C spectrum showing only $TeO_2$ remaining rather than a combination with Te.

Raman analysis of the degradation of $WTe_2$ follows a similar trend to $MoTe_2$, with minor changes up to 150 °C. Anneals above this temperature give spectra dominated by Te. There is some disparity between the Raman and XPS results of the 400 °C annealed film - XPS indicating the presence of $TeO_2$ while Raman indicates amorphous Te. This disparity may be due to the lower sampling depth of XPS compared to Raman spectroscopy.

As was the case for the selenides, the platinum TMD, $PtTe_2$ shows the highest thermal stability of the three telluride materials studied here. The Pt 4f core-level shows a small increase in the proportion of Pt metal in the film after 150 °C annealing. This indicates that either this is the primary route of degradation, or that the only UHV compatible product is through the formation of Pt metal rather than through oxide formation. After the 300 °C anneal, the $PtTe_2$ film has a substantial Pt metal signal equivalent to ~40% of measured Pt atoms. Despite this, the majority of the film is still $PtTe_2$ indicating significantly higher stability than the other telluride films that had no TMD detectable after this anneal. It is likely that 300 ˚C is near the upper thermal limit for $PtTe_2$ in ambient as the surface is fully Pt metal after the 400 °C anneal. The Te 3d core-level provides data consistent with the Pt 4f, showing a large amount of $PtTe_2$ is present up to the 300 °C anneal and only Pt metal and $TeO_2$ above this temperature.

As previously discussed, $PtTe_2$ has a less complex Raman spectrum compared to the other telluride TMDs and as the primary degradation route appears to be to elemental Pt this carries through to the post-anneal spectra. There are no new Raman peaks observed after any of the anneals but there is a noteworthy change in intensity after the 300 °C anneal with a large drop in the Raman signal from $PtTe_2$. This is likely due to thinning of the film and formation of Pt metal at the surface. Interestingly, there is a clear $PtTe_2$ Raman signal still apparent after 400 °C even though the XPS clearly shows the surface is entirely Pt metal and $TeO_2$. This result is similar to the results for the $MoS_2$ film, suggesting that after the 400 °C anneal the surface is fully oxidised but there is still $PtTe_2$ present beneath the surface layers.

**XPS Spectra of TMDs by Transition Metal**

Another valuable angle of comparison for this data is to compare the TMDs when grouped by their transition metal rather than their corresponding chalcogen. As these materials are all characterised with the same XPS metal core-levels, the data can be compared using just one core-level region.

**Molybdenum TMDs**

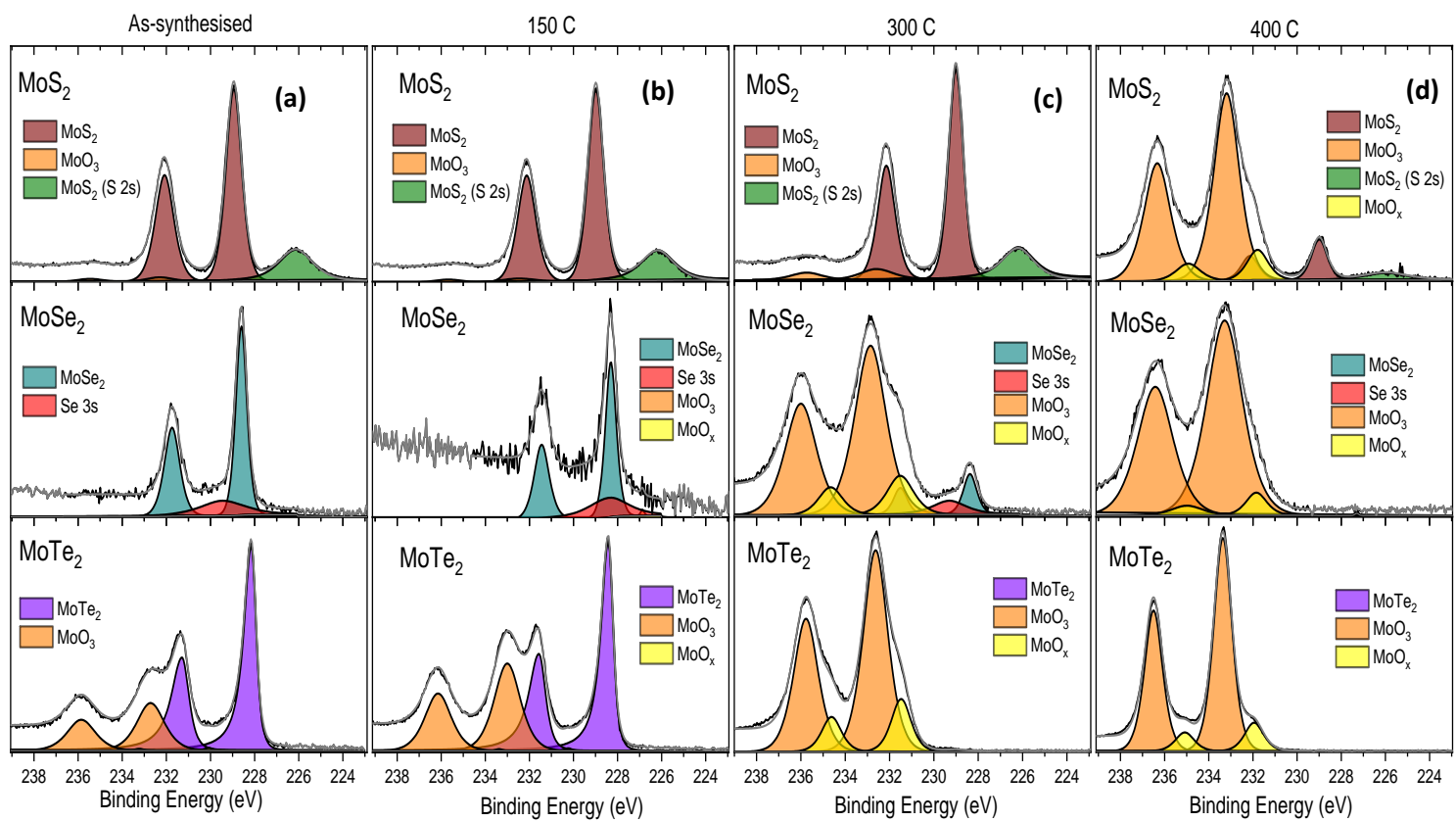

**Fig. 19** Mo 3d XPS spectra for the molybdenum TMDs showing multiple annealing steps. As-synthesised **(a)**, 150 °C **(b)**, 300 °C **(c)**, 400 °C **(d).**

Fig. 19 presents the Mo 3d region for $MoS_2$, $MoSe_2$, and $MoTe_2$. The difference in TMD peak position for the starting materials is clear here with a trend of decreasing peak binding energy position going sulfide > selenide > telluride. This is in-line with the trend in their electronegativities. $MoS_2$ and $MoSe_2$ have coincidentally overlapping core-levels with their respective chalcogens in the S 2s and Se 3s, complicating their spectra.

The difference in purity of the starting material is also clear, with $MoTe_2$ having ~30 % oxide while $MoS_2$ and $MoSe_2$ have negligible amounts. The relatively narrow FWHM of the $MoSe_2$ TMD peaks indicates a lower diversity of bonds and therefore a higher crystallinity than the other two materials, but the asymmetry of the $MoTe_2$ complicates comparison.

The 300 °C dataset displays the significant difference in stability between the three materials, with $MoS_2$ showing only a minor presence of $MoO_3$. $MoSe_2$ on the other hand is heavily oxidised on the surface, showing the formation of two oxides, a majority presence of MoO$_3$ as well as a minor oxide that it is not easily assigned. The MoTe$_2$ sample also has both of these oxides but no TMD is detected. This continues for the 400 °C anneal, with MoS$_2$ also now being predominantly oxidised showing the presence of both observed Mo oxides.

This analysis illustrates the relative thermal stability of these Mo TMDs. MoTe$_2$ has a significant amount of oxide in the starting material, which likely makes it more susceptible to further oxidation. These results present MoO$_3$ as the primary product in the ambient decomposition of Mo TMDs, a sub-oxide at ~232 eV is also formed in much lower amounts. A clear stability hierarchy of sulfide > selenide > telluride is also demonstrated for Mo TMD films.

**Tungsten TMDs**

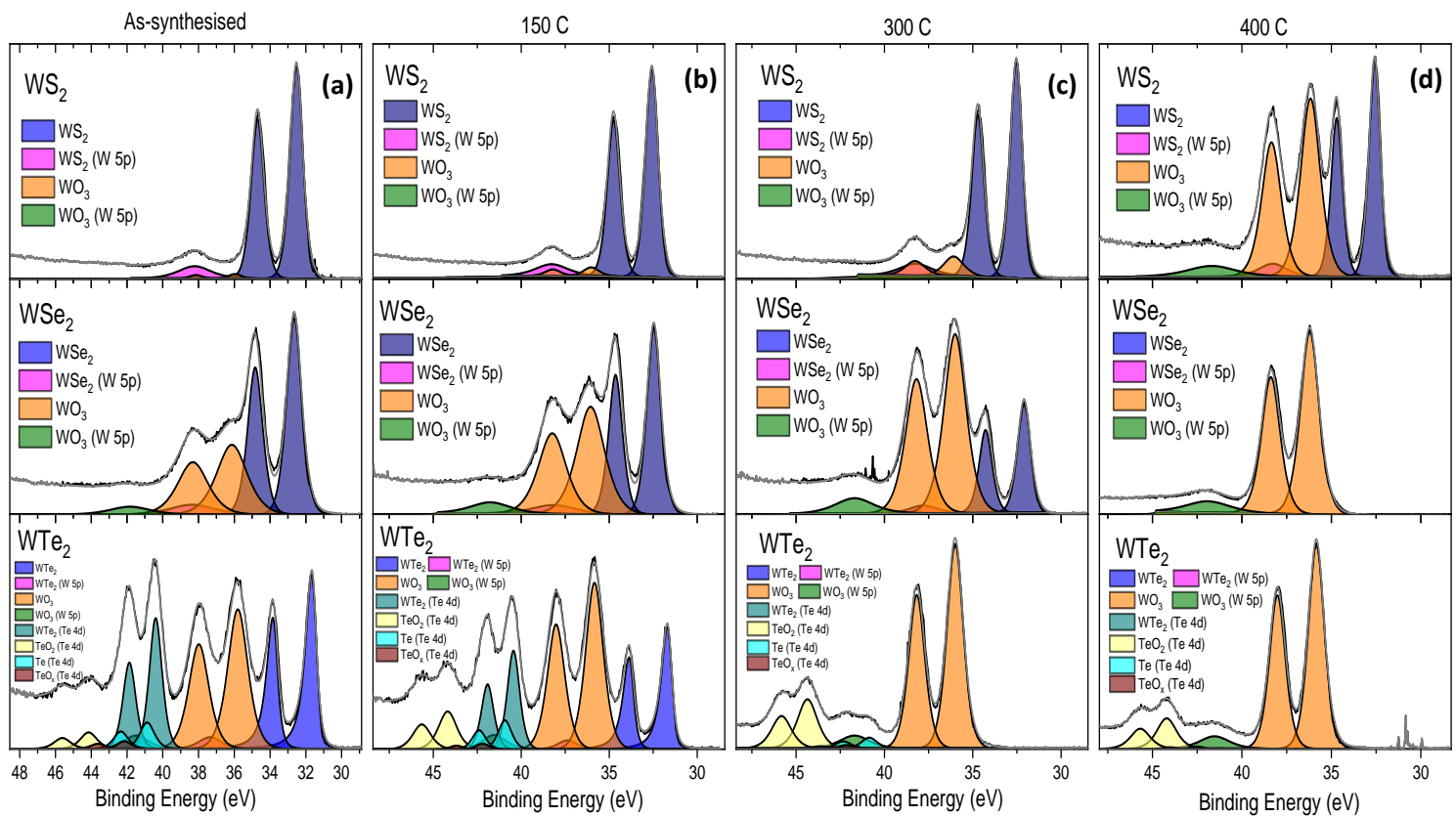

**Fig. 20** W 4f XPS spectra for the tungsten TMDs showing multiple annealing steps. As-synthesised **(a)**, 150 °C **(b)**, 300 °C **(c)**, 400 °C **(d).**

Fig.20 shows the W 4f core-level region of WS$_2$, WSe$_2$, and WTe$_2$. All three of the materials have WO$_3$ present in their as-synthesised state, albeit as a very minor component in the case of

WS2. The WTe2 spectrum, as previously discussed, is significantly more complex due to the presence of the overlapping Te 4d peaks. Unlike the Mo TMDs, the expected trend in peak position for the W TMDs is not as clear for the as-synthesised samples. This may be a result of charge-correction errors or a result of doping or charging from the oxide component in the films.

Analysis of the 300 ˚C annealed samples displays similar stability results to the Mo-based TMDs. The sulfide is only partially oxidised, while the selenide is heavily oxidised, and there is no TMD remaining on the WTe2 film surface. Notably, after the 400 ˚C anneal, the WS2 is still only 60% oxidised while the other materials are fully decomposed. For all three, WO3 is the only W oxide detected and appears to be the primary decomposition product.

**Platinum TMDs**

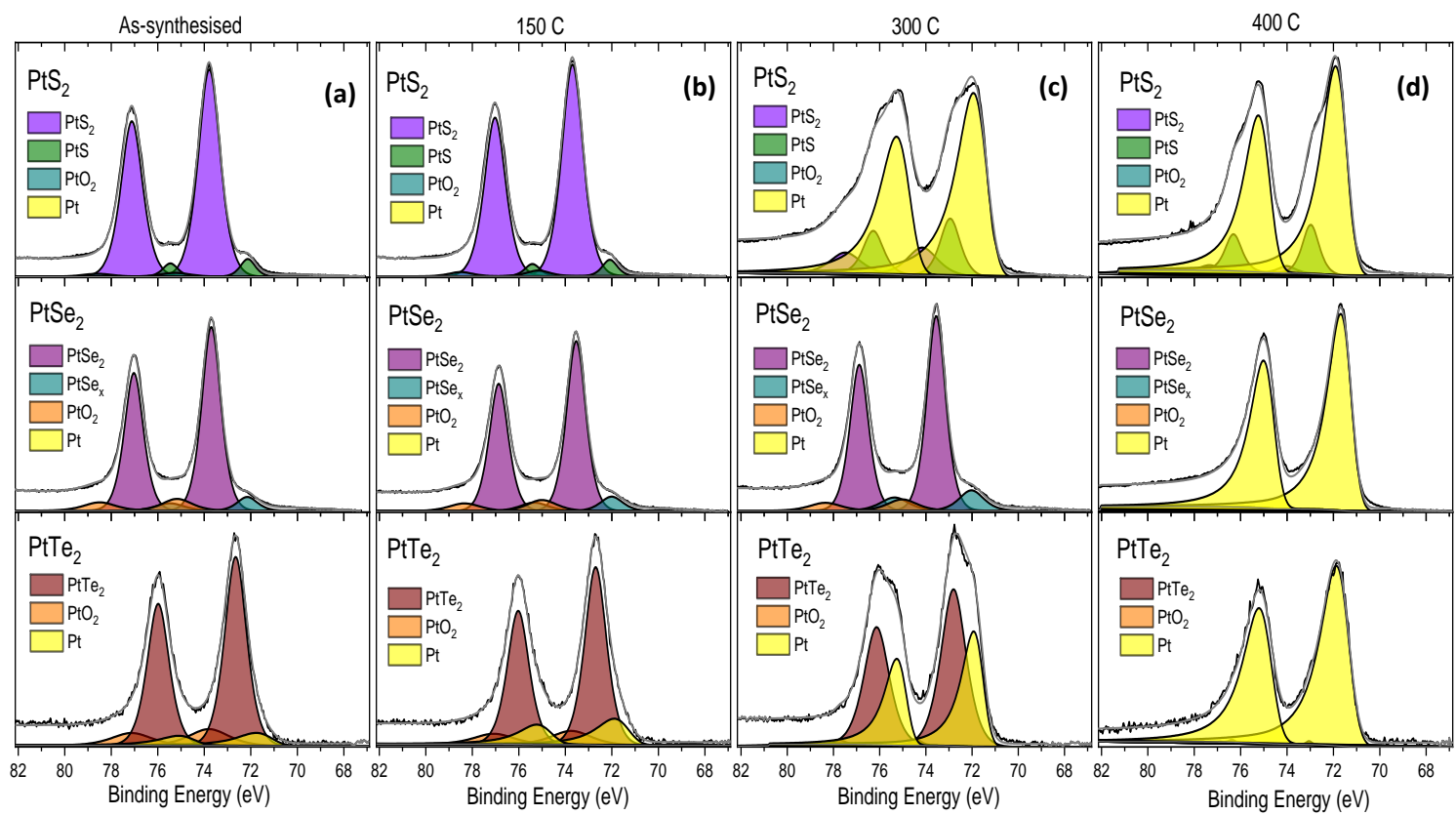

**Fig. 21** Pt 4f XPS spectra for the platinum TMDs showing multiple annealing steps. As-synthesised **(a)**, 150 °C **(b)**, 300 °C **(c)**, 400 °C **(d).**

The Pt 4f core-level regions of the three Pt TMDs, $PtS_2$, $PtSe_2$, and $PtTe_2$ are presented in Fig. 21. The binding energy position of the TMD peaks again aligns well with the relative electronegativities of the constituent chalcogens with sulfide having the highest binding energy and telluride the lowest. As previously discussed, the Pt 4f core-level region is less well understood than the equivalent regions for Mo and W TMDs and therefore the literature is underdeveloped in identification of the likely components for the various Pt TMDs.

All three Pt-based TMDs examined thus far have complex spectra with multiple components. The as-synthesised $PtTe_2$ is the only material found to also contain Pt metal, distinctive due to its asymmetric peak shape. Looking at the 300 °C annealed samples we can see that, in contrast to the Mo and W based TMDs, in this case $PtS_2$ is the least stable of the Pt TMDs. As previously mentioned, this is likely due to the relative stability of PtS. $PtSe_2$ shows marked stability amongst the Pt TMDs with negligible changes up to 300 ˚C. After 400 °C all of the Pt TMDs have decomposed, with $PtS_2$ maintaining a small PtS signal. It is noteworthy that all of the Pt TMDs have subsequently degraded to Pt metal instead of any oxide unlike the other TMDs studied. The rarity and low stability in ambient of Pt oxides is a known aspect in Pt catalysis.[118]

**Conclusions and perspectives**

A coherent methodology to examine and analyse the XPS spectra of a wide range of TMDs was discussed. Ten transition metal chalcogenides were synthesised through conversion of pre-deposited metal films on $SiO_2$. These included the sulfides, selenides, and tellurides of Mo, W, and Pt, together these encompass the most commonly studied materials in modern TMD literature. Raman spectroscopy, using multiple excitation wavelengths, was presented alongside the XPS analysis to complement, verify, and improve the characterisation of each material.

Consistency in interpretation was provided here in a clear, thorough, and broadly applicable manner. The parameters for XPS measurement and peak-fitting of each material were discussed individually, for each core-level.

The in-air thermal stability of the TMDs was examined after several ambient annealing points up to 400 °C. This allowed their gradual degradation to be observed, indicating their ambient stabilities. Importantly this annealing experiment provides a trend of evolving XPS and Raman spectra for each material, which can be used to greatly improve the accuracy of, and confidence in, the XPS analysis.

Beyond this, forced degradation of the TMD material allowed many of the most common features that may appear in the XPS spectra of the various TMDs to be illustrated and examined in detail. Furthermore, the essential benefits of complementary characterisation were clearly demonstrated with the combination of XPS and Raman providing a considerably more accurate understanding of the state of the material than either technique in isolation partly due to the differences in measurement depth and sensitivity.

The wide library of data assembled in this work allows the comparison of the degradation of the different films through the lens of either the transition metal or the chalcogen. Accordingly, it can be qualitatively concluded that the stability of the TMDs follows the general trend of sulfides > selenides > tellurides. Analogously it appears the stability of TMDs grouped by transition metal is platinum > tungsten > molybdenum. A standout exception to this is $PtS_2$, as one of the least stable TMDs regarding ambient annealing, this is likely due to the higher stability of the monosulfide in this case.

Some common limitations, open questions, and other continuing issues involved with XPS and Raman analysis of the 10 materials featured here were discussed to raise awareness and to convey the nuanced approach required. This includes understanding that the spectra and interpretations provided here are a much-needed guideline but are not an exhaustive manual.

It is envisioned that this library of data will act as a guideline to enable researchers of all levels of expertise to characterise, compare, and discuss the XPS and Raman spectra of TMD materials in a more accessible manner. This is particularly worthwhile for those involved in applications-focused research where the characterisation of the TMD materials' composition may not be the primary focus.

**Acknowledgements**

This work was supported by Science Foundation Ireland (SFI) through (15/IA/3131, 12/RC/2278_P2, 15/SIRG/3329). G.S.D and O.H. acknowledge the European Commission under the project Queformal [829035] and the German Ministry of Education and Research (BMBF) under the projects ACDC [13N15100] and NobleNEMS [16ES1121].

# Supplementary Information

# Synthesis and thermal stability of TMD thin films: A comprehensive XPS and Raman study

*Conor P. Cullen[1,2], Oliver Hartwig[3], Cormac Ó Coileáin [1,2], John B. McManus[1,2], Lisanne Peters[1,2], Cansu Ilhan[1,2], Georg S. Duesberg[1,2,3], Niall McEvoy[1,2]*

[1] School of Chemistry, Trinity College Dublin, Dublin 2, D02 PN40, Ireland

[2] AMBER Centre, CRANN Institute, Trinity College Dublin, Dublin 2, Ireland

[3] Institute of Physics, EIT 2, Faculty of Electrical Engineering and Information Technology, Universität der Bundeswehr, 85579 Neubiberg, Germany

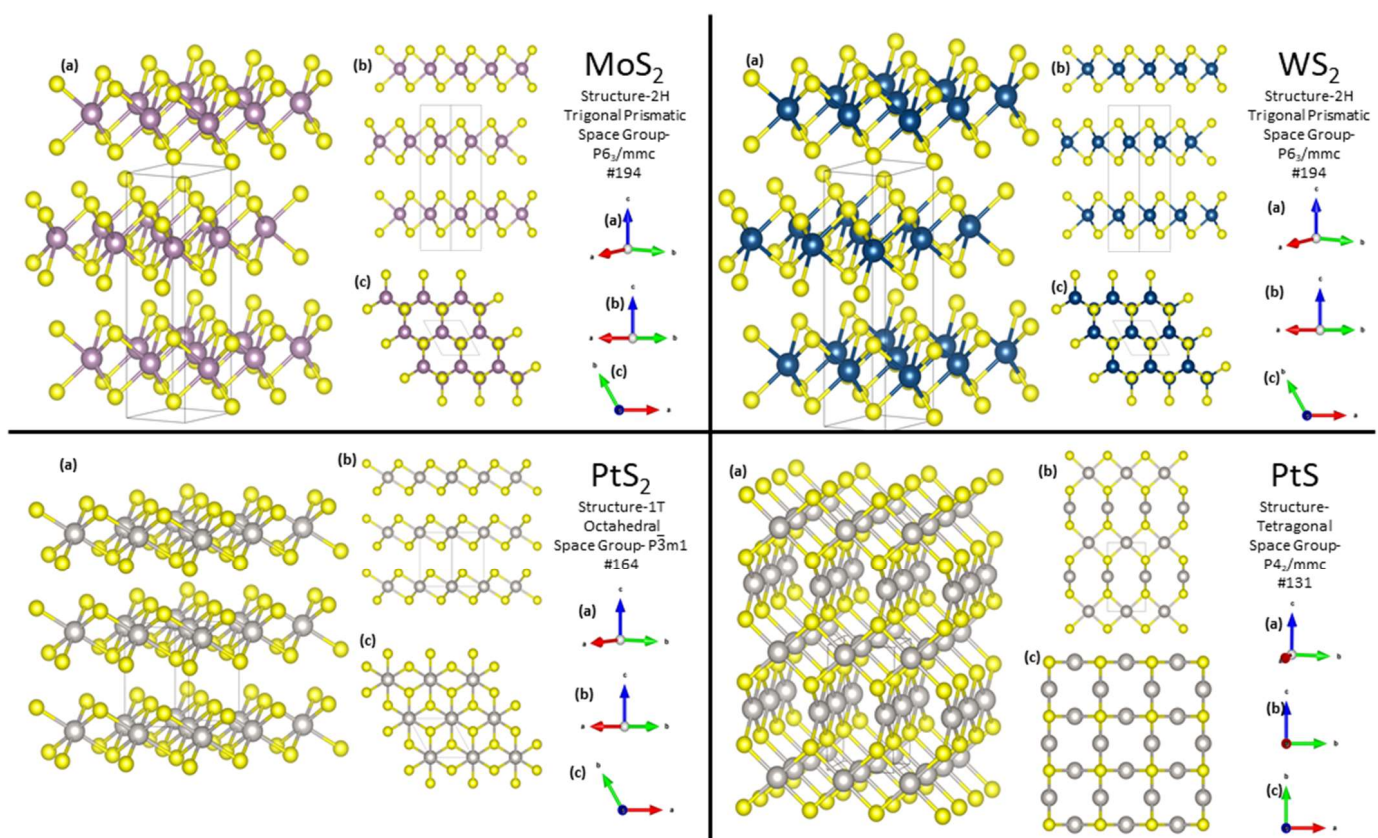

**Fig. S1** Structural figures for $MoS_2$, $WS_2$, $PtS_2$, and PtS.

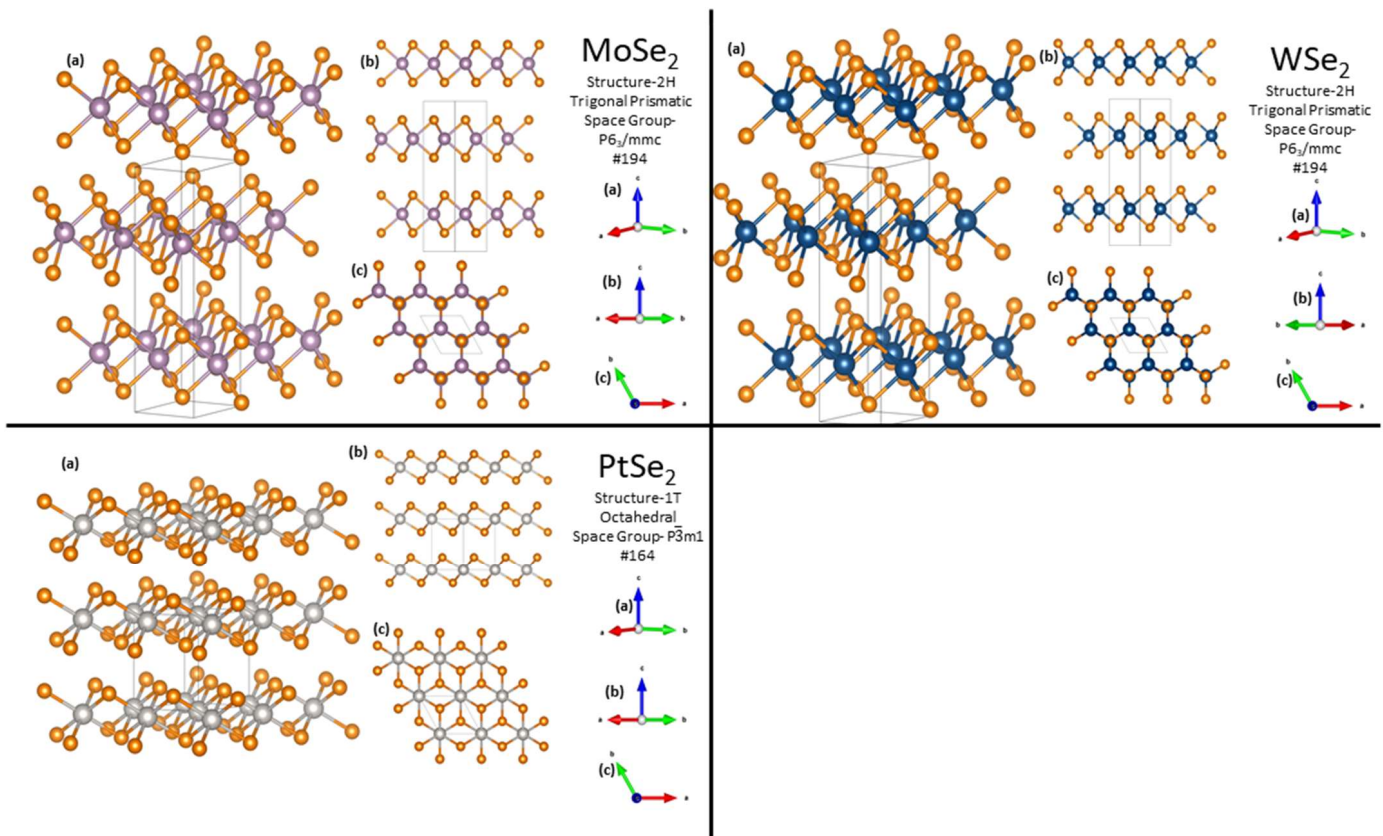

**Fig. S2** Structural figures for MoSe$_2$, WSe$_2$, PtSe$_2$

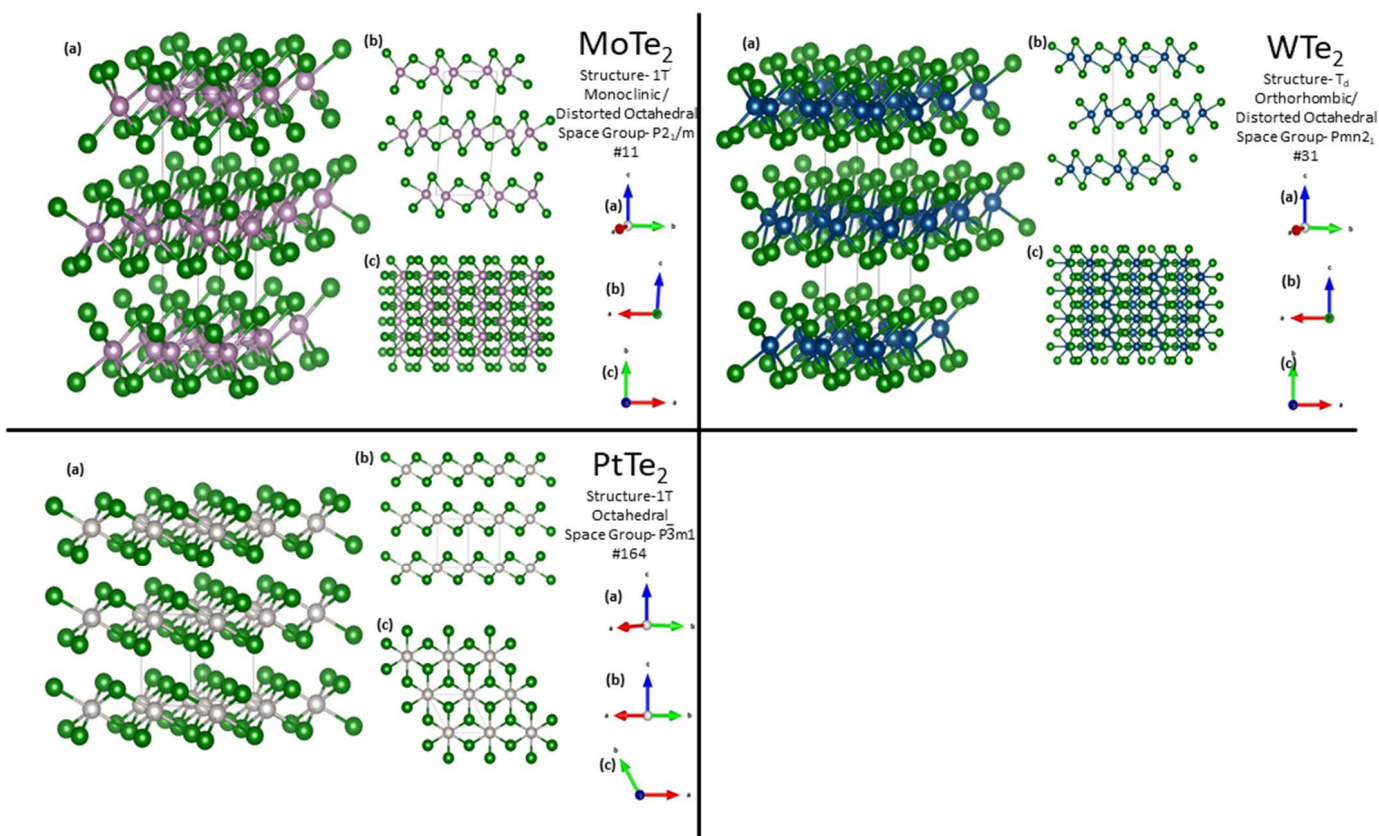

**Fig. S3** Structural figures for MoTe$_2$, WTe$_2$, PtTe$_2$

**Table.S1** Details of the XPS fitting for MoS$_2$ core-levels

| Mo 3d | MoS$_2$ | MoS$_2$ (S 2s) | MoO$_3$ | S 2p | MoS$_2$ |
|---|---|---|---|---|---|
| Position (eV) | 228.94 | 226.12 | 232.32 | Position (eV) | 161.84 |
| Peak shape | GL(60) | SGL(60) | GL(60) | Peak shape | GL(60) |
| FWHM (eV) | 0.83 | 1.84 | 1.46 | FWHM (eV) | 0.81 |
| Mo % | 96.1 | - | 3.9 | S % | 100 |

**Table S2** Details of the XPS fitting for WS$_2$ core-levels

| W 4f | WS$_2$ | WS$_2$ (W 5p) | WO$_3$ | WO$_3$ (W 5p) | S 2p | WS$_2$ |
|---|---|---|---|---|---|---|
| Position (eV) | 32.52 | 38.22 | 35.96 | 41.69 | Position (eV) | 162.02 |
| Peak shape | GL(60) | GL(60) | GL(60) | GL(60) | Peak shape | GL(60) |
| FWHM (eV) | 0.83 | 2.15 | 0.82 | 3.5 | FWHM (eV) | 0.81 |
| W % | 98 | - | 2 | - | S % | 100 |

**Table. S3** Details of the XPS fitting for $PtS_2$ core-levels

| Pt 4f | $PtS_2$ ($Pt^{4+}$) | PtS ($Pt^{2+}$) | $PtO_2$ | S 2p | $PtS_2$ ($S^{2-}$) | PtS ($S^{2-}$) | Elemental S | $S^{4+}$ |
|---|---|---|---|---|---|---|---|---|
| Position (eV) | 73.79 | 72.13 | 75.26 | Position (eV) | 162.18 | 162.64 | 163.78 | 166.65 |
| Peak shape | GL(60) | GL(60) | GL(60) | Peak shape | GL(60) | GL(60) | GL(60) | GL(60) |
| FWHM (eV) | 1.09 | 0.76 | 1.50 | FWHM (eV) | 0.93 | 0.72 | 2 | 2.35 |
| Pt % | 92.1 | 5.5 | 2.4 | S % | 84.7 | 3 | 5.3 | 7 |

**Table S4** Details of the XPS fitting for PtS core-levels

| Pt 4f | PtS ($Pt^{2+}$) | $PtS_2$ ($Pt^{4+}$) | $PtO_2$ | S 2p | PtS ($S^{2-}$) | $PtS_2$ ($S^{2-}$) | Elemental S | $S^{4+}$ |
|---|---|---|---|---|---|---|---|---|
| Position (eV) | 72.05 | 73.33 | - | Position (eV) | 162.53 | 161.33 | 163.95 | 166.83 |
| Peak shape | GL(60) | GL(60) | - | Peak shape | GL(60) | GL(60) | GL(60) | GL(60) |
| FWHM (eV) | 0.92 | 1.36 | - | FWHM (eV) | 0.81 | 0.7 | 1.15 | 2.5 |
| Pt % | 96 | 4 | - | S % | 76.9 | 5 | 10.6 | 7.5 |

**Table S5** Details of the XPS fitting for MoSe$_2$ core-levels

| Mo 3d | MoSe$_2$ | MoSe$_2$ (Se 3s) | Se 3d | MoSe$_2$ | MoSe$_x$ |
|---|---|---|---|---|---|
| Position (eV) | 228.6 | 229.44 | Position (eV) | 54.34 | 54.61 |
| Peak shape | GL(60) | SGL(60) | Peak shape | GL(60) | GL(60) |
| FWHM (eV) | 0.92 | 2.4 | FWHM (eV) | 0.6 | 0.71 |
| Mo % | 100 | - | Se % | 80.5 | 16.5 |

**Table S6** Details of the XPS fitting for WSe$_2$ core-levels

| W 4f | WSe$_2$ | WSe$_2$ (W 5p) | WO$_3$ | WO$_3$ (W 5p) | Se 3d | WSe$_2$ | WSe$_x$ |
|---|---|---|---|---|---|---|---|
| Position (eV) | 32.65 | 38.35 | 36.13 | 41.83 | Position (eV) | 54.8 | 55.32 |
| Peak shape | GL(60) | GL(60) | GL(60) | GL(60) | Peak shape | GL(60) | GL(60) |
| FWHM (eV) | 0.99 | 2.9 | 2 | 2.62 | FWHM (eV) | 0.85 | 0.85 |
| W % | 58.15 | - | 41.85 | - | Se % | 79 | 21 |

**Table S7** Details of the XPS fitting for PtSe$_2$ core-levels

| Pt 4f | PtSe$_2$ (Pt$^{4+}$) | PtSe$_x$ (Pt$^{2+}$) | PtO$_2$ | Se 3d | PtSe$_2$ | PtSe$_x$ | Se oxide | Pt 5p |
|---|---|---|---|---|---|---|---|---|
| Position (eV) | 73.7 | 72.15 | 75.17 | Position (eV) | 55.03 | 55.77 | 58.33 | 53.67 |
| Peak shape | GL(60) | GL(60) | GL(60) | Peak shape | GL(60) | GL(60) | GL(60) | GL(60) |

| FWHM (eV) | 0.96 | 1.02 | 1.5 | FWHM (eV) | 0.77 | 0.82 | 2.5 | 2.62 |
|---|---|---|---|---|---|---|---|---|
| Pt % | 84.6 | 6.8 | 8.6 | Se % | 73.4 | 18.6 | 8 | - |

Table S8 Details of the XPS fitting for MoTe$_2$ core-levels

| Mo 3d | MoTe$_2$ | MoO$_3$ | Te 3d | MoTe$_2$ | TeO$_2$ | Elemental Te | TeO$_x$ |
|---|---|---|---|---|---|---|---|
| Position (eV) | 228.15 | 232.71 | Position (eV) | 572.47 | 576.26 | 573.18 | 574.1 |
| Peak shape | LA(1.1, 2.3, 2) | GL(60) | Peak shape | GL(60) | GL(60) | GL(60) | GL(60) |
| FWHM (eV) | 0.52 | 1.46 | FWHM (eV) | 0.95 | 0.139 | 1.10 | 1.44 |
| Mo % | 70.1 | 29.9 | Te % | 58.5 | 22.1 | 13.6 | 5.8 |

Table S9 (a) Details of the XPS fitting for the W 4f core-level of WTe$_2$

| W 4f | WTe$_2$ | WTe$_2$ (W 5p) | WO$_3$ | WO$_3$ (W 5p) | WTe$_2$ (Te 4d) | TeO$_2$ (Te 4d) | Te (Te 4d) | TeO$_x$ (Te 4d) |
|---|---|---|---|---|---|---|---|---|
| Position (eV) | 31.65 | 37.35 | 35.80 | 41.50 | 40.40 | 44.15 | 40.87 | 42.15 |
| Peak shape | LA(1.3,2.5,11) | LA(1.3,2.5,11) | GL(60) | GL(60) | GL(60) | GL(60) | GL(60) | GL(60) |
| FWHM (eV) | 0.61 | 1.48 | 1.33 | 1.64 | 0.84 | 1.08 | 0.95 | 0.95 |
| W% | 44.2 | - | 55.8 | - | - | - | - | - |

**Table S9 (b)** Details of the XPS fitting for the Te 3d core-level of WTe$_2$

| Te 3d | WTe$_2$ | TeO$_2$ | Elemental Te | TeO$_x$ |
|---|---|---|---|---|
| Position (eV) | 572.39 | 576.18 | 573.03 | 574.14 |
| Peak shape | GL(60) | GL(60) | GL(60) | GL(60) |
| FWHM (eV) | 0.98 | 1.34 | 1.64 | 1.17 |
| Te % | 63.7 | 18 | 16 | 2.3 |

**Table S10** Details of the XPS fitting for PtTe$_2$ core-levels

| Pt 4f | PtTe$_2$ | Elemental Pt | PtO$_2$ | Te 3d | PtTe$_2$ | TeO$_2$ | Elemental Te | TeO$_x$ |
|---|---|---|---|---|---|---|---|---|
| Position (eV) | 72.65 | 71.30 | 73.80 | Position (eV) | 572.92 | 575.57 | 573.36 | 573.92 |
| Peak shape | GL(60) | LA(1.2,85,70) | GL(60) | Peak shape | GL(60) | GL(60) | GL(60) | GL(60) |
| FWHM (eV) | 1.11 | 1.5 | 2 | FWHM (eV) | 0.95 | 1.59 | 1.80 | 0.97 |
| Pt % | 77.7 | 10 | 12.3 | Te % | 50.8 | 35 | 10.1 | 4.1 |

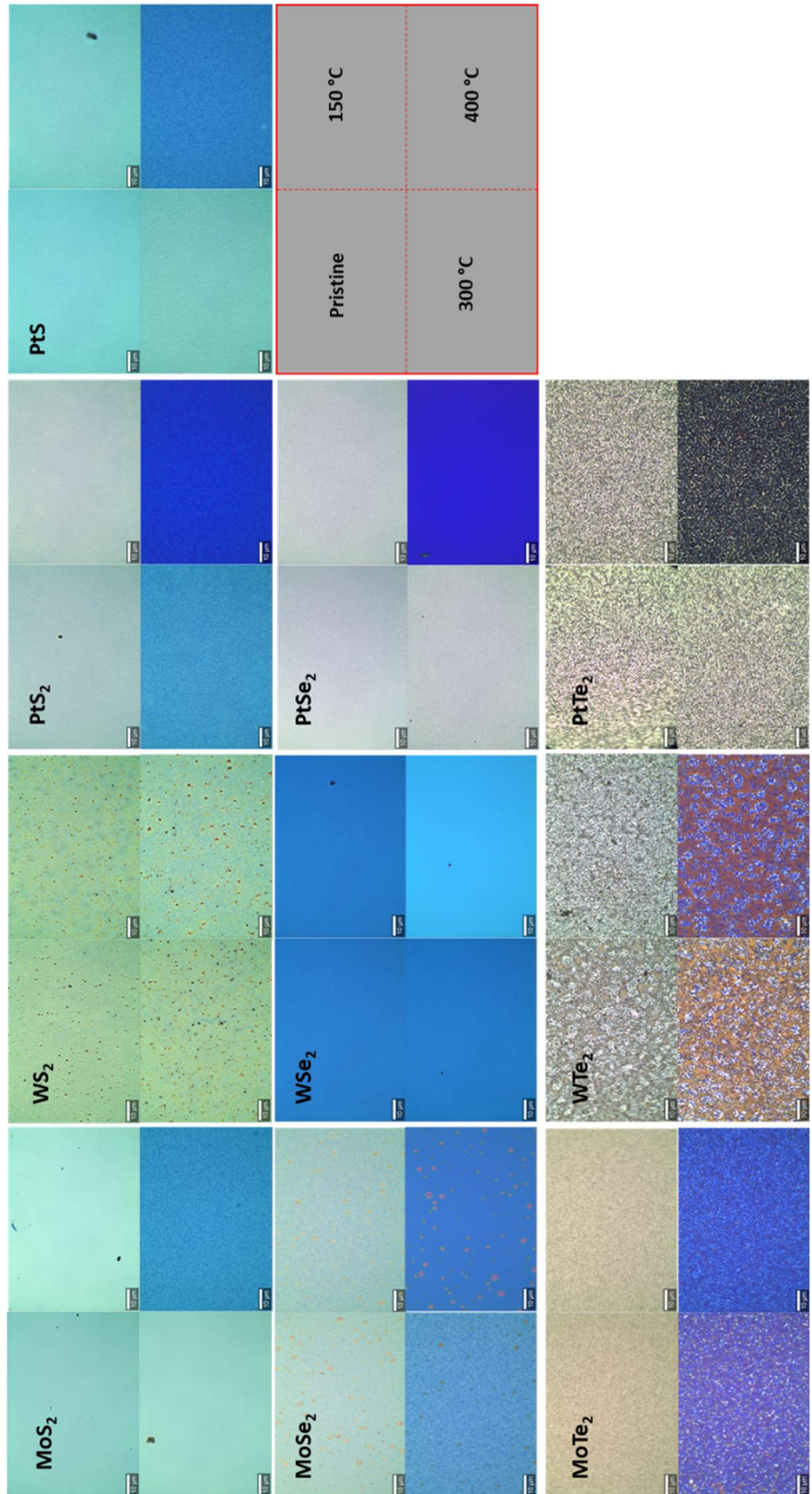

**Fig.S4** Optical micorscope images of the films after the various anneals

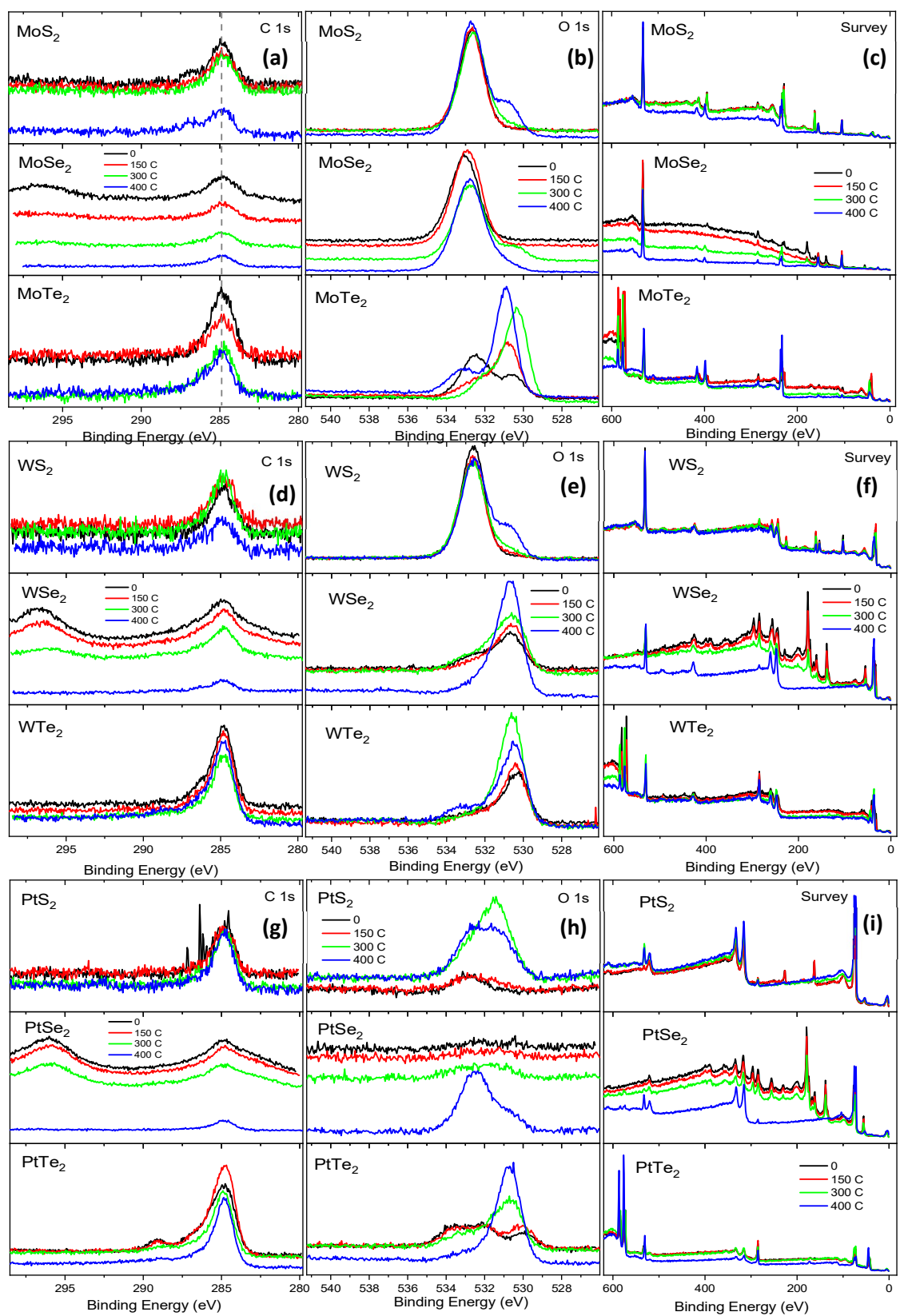

**Fig.S5** The accompanying XPS spectra for the TMDs measured here at each anneal temperature. C 1s, O 1s, and survey XPS data for the Mo TMDs **(a–c)**, W TMDs **(d–f)**, and Pt TMDs **(g–i)**

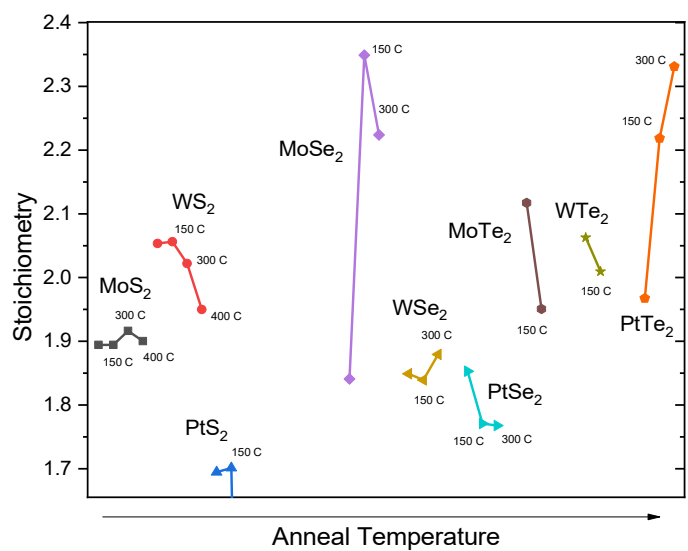

**Fig.S6** Calculated stoichiometry data from the collection of samples

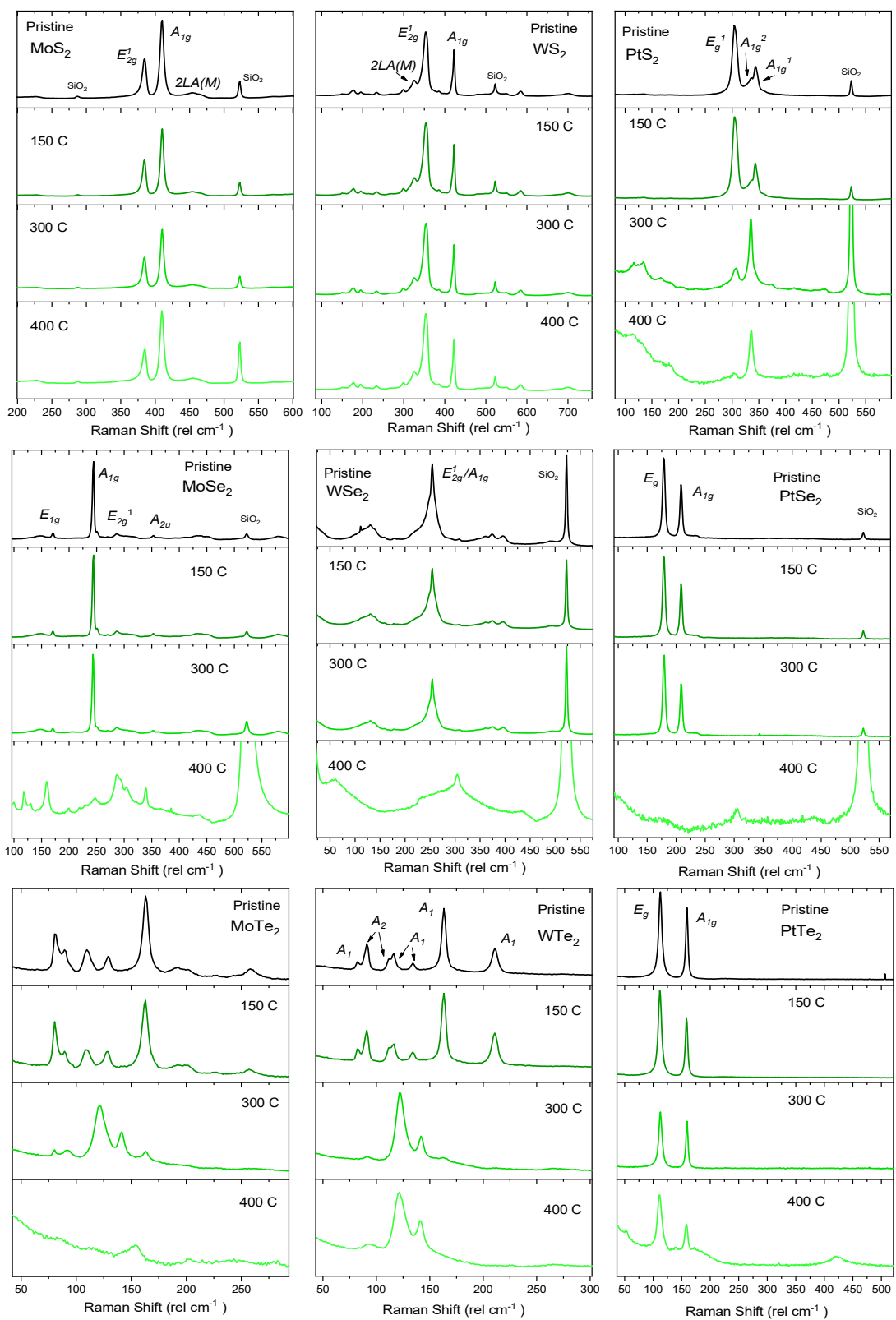

**Fig.S7** 532 nm excitation Raman spectra of the nine TMD films pre- and post-anneal at three temperatures.

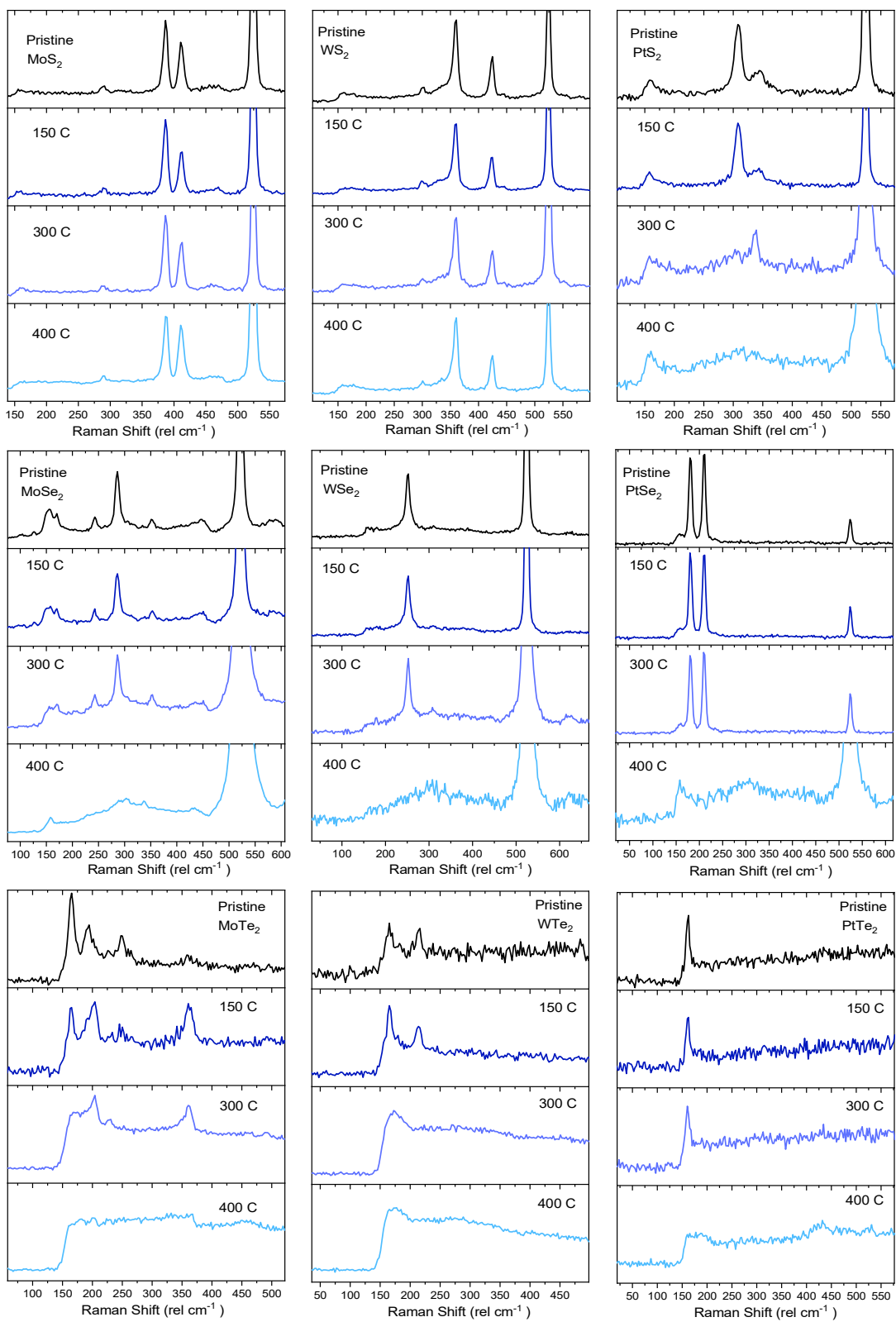

**Fig.S8** 405 nm excitation Raman spectra of the nine TMD films pre- and post-anneal at three temperatures.

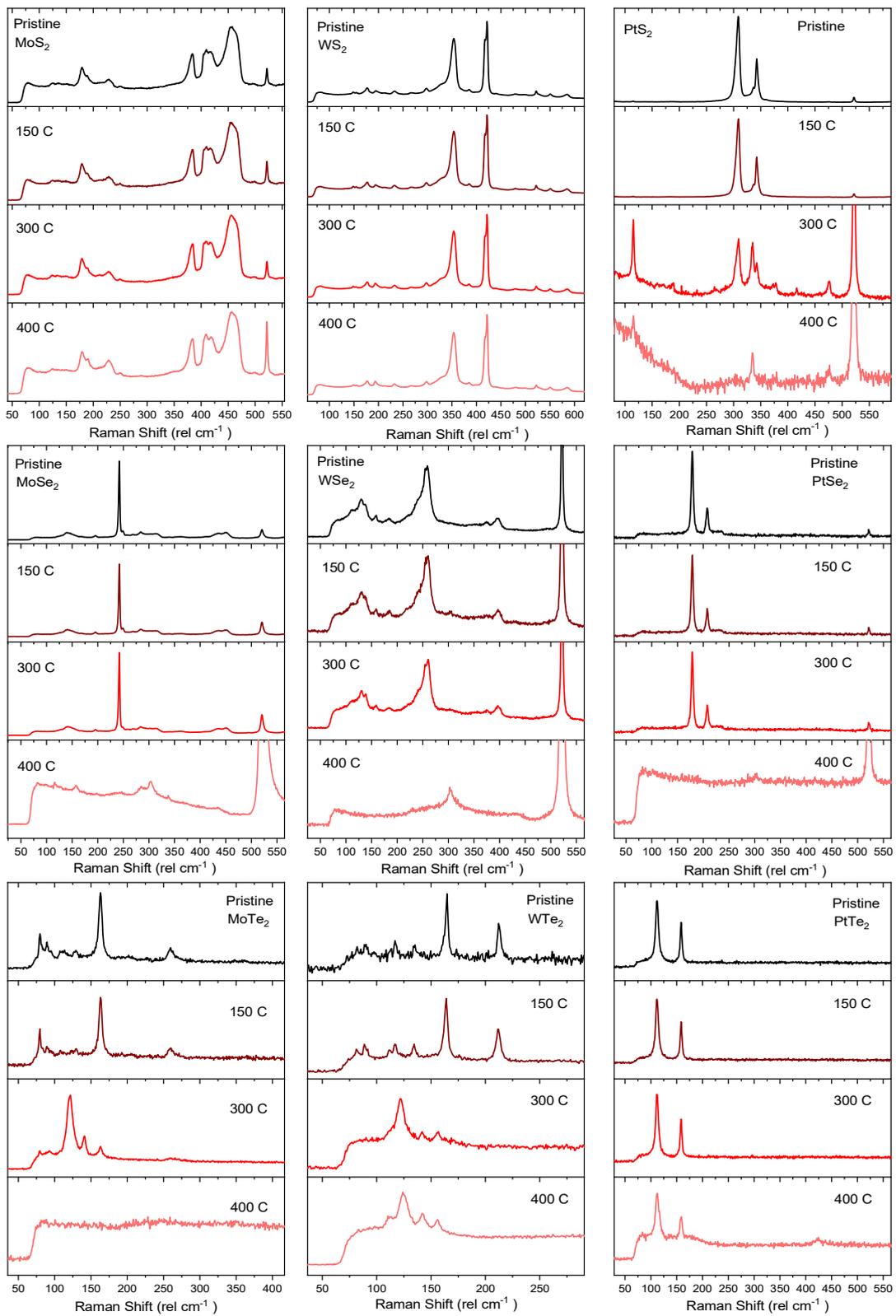

**Fig.S9** 633 nm excitation Raman spectra of the nine TMD films pre- and post-anneal at three temperatures.

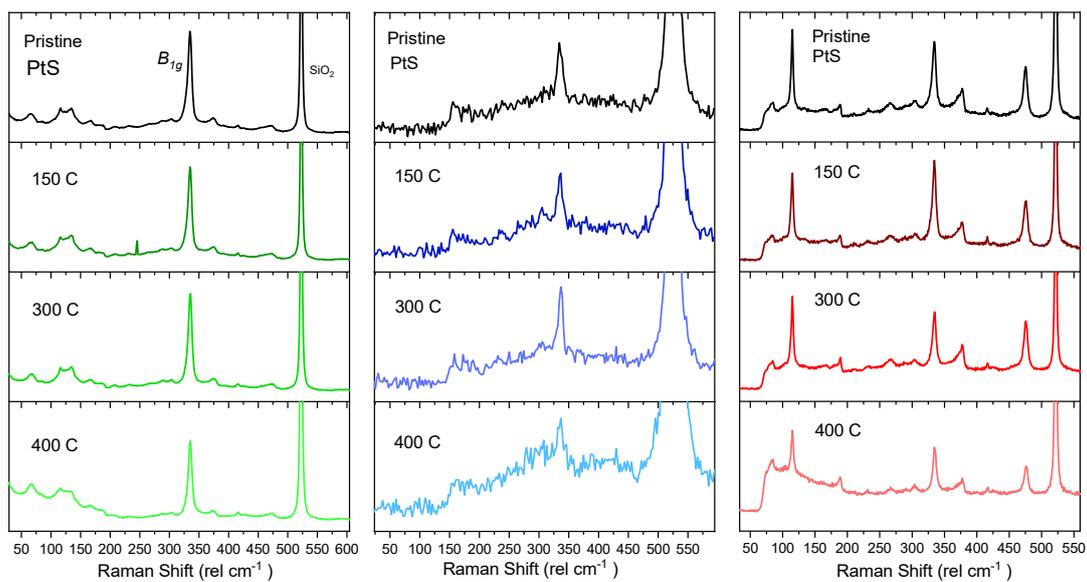

**Fig.S10** Raman spectra of PtS with 532, 405, and 633 nm excitation energy, pre- and post-anneal at three temperatures.